\documentclass[conference]{IEEEtran}
\AtBeginDocument{%
	\providecommand\BibTeX{{%
			\normalfont B\kern-0.5em{\scshape i\kern-0.25em b}\kern-0.8em\TeX}}}

\usepackage[utf8]{inputenc}
\usepackage{makecell, longtable} 
\usepackage{arydshln} 
\usepackage{multirow}
\usepackage{pifont} 
\usepackage{array}
\usepackage{graphicx}
\usepackage{wrapfig}
\usepackage{paralist}
\usepackage{underscore}
\usepackage[english]{babel}
\usepackage{multicol}        
\usepackage[bottom]{footmisc} 
\usepackage{float}
\usepackage{xcolor}

\usepackage{tabularx,booktabs}
\newcolumntype{Y}[1]{>{\raggedright\let\newline\\\arraybackslash\hspace{0pt}}m{#1}} 

\newcolumntype{C}[1]{>{\centering\let\newline\\\arraybackslash\hspace{0pt}}m{#1}}

\usepackage{tikz}
\newcommand{\cmark}{\ding{51}}
\newcommand{\xmark}{\ding{55}}
 
\ifCLASSINFOpdf
\else
\fi

\usepackage[hyphens]{url}

\author{\IEEEauthorblockN{Mohammad Hossein Tabatabaei}
	\IEEEauthorblockA{Department of Informatics\\
			University of Oslo\\
			mohammht@ifi.uio.no}
	\and
	\IEEEauthorblockN{Roman Vitenberg}
	\IEEEauthorblockA{Department of Informatics\\
			University of Oslo\\
		    romanvi@ifi.uio.no}
	\and
	\IEEEauthorblockN{Narasimha Raghavan Veeraragavan}
	\IEEEauthorblockA{Department of Informatics\\
		University of Oslo\\
		raghavan@ifi.uio.no}
	}

\begin{document}
%
\title{Understanding blockchain: definitions, architecture, design, and system comparison}
\maketitle

\begin{abstract}
The explosive advent of the blockchain technology has led to hundreds of blockchain systems in the industry, thousands of academic papers published over the last few years, and an even larger number of new initiatives and projects. Despite the emerging consolidation efforts, the area remains highly turbulent without systematization, educational materials, or cross-system comparative analysis.

In this paper, we provide a systematic and comprehensive study of four popular yet widely different blockchain systems: Bitcoin, Ethereum, Hyperledger Fabric, and IOTA. The study is presented as a cross-system comparison, which is organized by clearly identified aspects: definitions, roles of the participants, entities, and the characteristics and design of each of the commonly used layers in the cross-system blockchain architecture. Our exploration goes deeper compared to what is currently available in academic surveys and tutorials. For example, we provide the first extensive coverage of the storage layer in Ethereum and the most comprehensive explanation of the consensus protocol in IOTA. The exposition is due to the consolidation of fragmented information gathered from white and yellow papers, academic publications, blogs, developer documentation, communication with the developers, as well as additional analysis gleaned from the source code. We hope that this survey will help the readers gain in-depth understanding of the design principles behind blockchain systems and contribute towards systematization of the area.
\end{abstract}


%
\IEEEpeerreviewmaketitle

\section{Introduction}
\label{sec:introduction}


Since Bitcoin inception and deployment back in 2009, it has been creating waves of discussions and debates about its success as the first widely popular cryptocurrency effort.
Bitcoin did away with a broker entity and showed that it is fundamentally possible to have a working cooperative economic model in practice where the technology does all the mediation. 

In 2014--2016, there occurred an explosive extension in the range of applications, with everybody talking about the \emph{Blockchain Paradigm} underlying the technology that enables cooperative economic models and cooperative storage and management of agreements. 
According to~\cite{coindesk-research}, the annual revenue of blockchain-based enterprise applications worldwide will reach \$19.9 billion by 2025, up from about \$2.5 billion in 2016. 

Declared intention of future use of blockchain in cooperative healthcare provision is making global news~\cite{blockchain-healthcare}. 
Cooperative energy markets based on microgrids are viewing blockchain as a major technological enabler~\cite{blockchain-energy}. 
Blockchain is becoming a common element in solutions for decentralized digital identity (DID)~\cite{blockchain-did}, certificate storage~\cite{blockchain-certificates}, land registries~\cite{blockchain-land}, and supply chains~\cite{blockchain-supply}.
Microsoft declares ``blockchain'' as one of the key ``must win'' technologies for their Azure platform and business~\cite{ms-blockchain}. Similarly, IBM unveils a new ambitious blockchain service and strategy~\cite{ibm-blockchain}. A long row of major enterprises such as Accenture, Cisco, Citibank, Facebook, Disney Studios, Goldman Sachs, and HSBC have publicly declared their investment in the technology with the intent of exploring the potential for exploitation, which has resulted in an establishment of industrial blockchain associations such as Diem~\cite{diem}.
Governments in North America, Europe, and Asia are advancing blockchain-related strategy and legislation, reflecting their significant interest towards the advent and utilization of the technology~\cite{UK-document, Netherlands-strategy, ebp, ebsi, China-document,estonia-blockchain}.
Blockchain has become one of the hottest topics at many societal, industrial, and academic conferences.

The above has resulted in hundreds of blockchain systems in the industry, thousands of academic papers published over the last few years, and an even larger number of new initiatives and projects. Following this explosive development, the area currently remains tumultuous, without commonly accepted terminology and with occasionally diverging concepts and ideas. For example, we list five different definitions of the term "blockchain" in Section~\ref{sec:reflection}, all being used in the literature and in description of actual systems.

Emerging surveys such as~\cite{ellervee2017comprehensive} and~\cite{belotti2019vademecum} among all have made an important first step towards orderly understanding of the area. However, most of the published surveys are focusing on one system (such as Bitcoin~\cite{bonneau2015sok, tschorsch2016bitcoin}), a particular class of blockchain (e.g., DAG-based~\cite{sok-dag}), or an individual aspect of blockchain (such as proof-of-X or defense mechanisms against security attacks of a specific type). In-depth comparisons across systems of different type  are exceedingly rare, and so are standardization initiatives or attempts to establish general taxonomies. Thus, despite the consolidation efforts, the area remains highly turbulent without systematization. Accordingly, there are no textbooks or comprehensive educational materials, except for Bitcoin itself~\cite{bitcoinText}.

Today, the only way to gain in-depth technical understanding of the design behind almost any blockchain system is to read typically outdated white and yellow papers followed by perusing the continuously updated technical documentation, followed by scanning hundreds of blog posts by the developers and conflicting forum posts by the users and finally, by studying the source code.

\subsection{Contributions}

Our main contribution is the first systematic and comprehensive comparative study of blockchain design across different systems. While many existing works provide a focused cross-system comparison on a particular aspect, we aim to provide a comparative design insight that goes far beyond the information that is currently available in the literature. The comparison is performed across four blockchain systems representative of different blockchain strands (i.e., Bitcoin, Ethereum, Hyperledger Fabric, and IOTA) with divergence in the design priorities, architectural elements, performance, and even roles of the participants and basic definitions.
While IOTA is not as popular as the other three, it is the most commonly used DAG-based system and it was ranked at the fourth place~\cite{marketCap} by the market cap in 2017.


First, we present generic roles of the participants in blockchain along with the significance of each role and explain how the entities (i.e., computing devices of different categories) in a blockchain system of each type map to these roles. We introduce a generic layered architecture that applies to all blockchain systems regardless of the type. The study of the four systems is organized across these layers so that the design of each layer is considered separately from the rest. This methodology allows us to conduct a comprehensive cross-system comparison. The comparison is organized by clearly identified aspects: definitions, roles, entities, and the characteristics and design of each of the layers. We also contrast the performance of the four systems based on the previously published information and explain the reasons for the differences.

Our exploration goes deeper compared to what is currently available in academic surveys and tutorials. For example, we provide the first extensive coverage of the storage layer in Ethereum and the most comprehensive explanation of the consensus protocol in IOTA. The exposition is due to the consolidation of fragmented information about popular systems from blogs, developer documentation, and studying the source code.

Our main emphasis is on the education and pedagogical exposition that lends itself to courses and tutorials. While such descriptions exist for Bitcoin, no such materials are available for Ethereum and IOTA to the best of our knowledge. A more fine-grained summary of our contribution along with the comparison with existing surveys is available in Section~\ref{sec:relatedWorks}.


Our study focuses on the design of the systems themselves rather than on application mechanisms developed atop them. In particular, we do not cover hybrid storage systems that combine on-chain and off-chain elements. Besides, the scope does not include functional features such as sharding that are still in development and that are not supported by the current versions of the blockchain systems.

\subsection{Roadmap}

In Section~\ref{sec:understanding}, we first contrast various blockchain definitions and reflect on the discrepancies in the commonly used terminology. Then, we introduce a list of roles of the participants and a layered blockchain architecture that are both applicable to all blockchain systems regardless of the type. In Section~\ref{sec:representative}, we present an overview of the four systems, discuss the entities in each, and explain how they map to the generic roles. In Section~\ref{sec:comparelayers}, we provide a layer-by-layer comparison between the four systems while covering a variety of design aspects and characteristics. In Section~\ref{sec:relatedWorks}, we contrast our work with other state-of-the-art surveys. Finally, we present our conclusions in Section~\ref{sec:conclusions}.

\section{Understanding Blockchain: Definitions and Concepts}
\label{sec:understanding}

In this section, we review and reflect upon central definitions and concepts of blockchain technologies.

\subsection{Reflection on Various Blockchain Definitions}
\label{sec:reflection}

The term of ``blockchain'' generally refers to a paradigm for maintaining information in a distributed system that is characterized by a number of properties. Since there is no specification or established standards in \the\year\ yet, different concretizations of this general definition have been adopted in the literature and existing popular blockchain-based systems. 

Distributed Ledger Technologies (DLTs, in short) is a well-defined term: it refers to a system that records a ledger of transactions or a history of changes to the system state. The ledger is usually hard to tamper with, which is a boon for security, yet it also makes it hard to perform desirable changes, e.g., to prune the history or compact the ledger.

While people tend to equate blockchain with DLTs, both narrower and broader meanings of ``blockchain'' are in use. Literally, blockchain means ``a chain of blocks'', which implies a specific data structure for the ledger implementation. A chain of blocks precludes any parallelism between the transactions, however, which has a negative impact on the performance. Some ledger implementations use more a complex data structure such as braids~\cite{braids} or a directed acyclic graph (DAG) in IOTA, which allow some degree of parallelism by retaining concurrently proposed competing blocks and merging them. Since the term of DLT does not imply any specific data structure, it covers such a generalization. On the other hand, the term of ``blockchain'' becomes a misnomer in that case.
In absence of more refined terminology today, ``blockchain'' is used in the literature to refer to a chain of blocks or generalized DLTs.

To add to the confusion, some systems in this domain do not maintain a distributed ledger at all. For example, Corda~\cite{corda} allows participating computing devices to agree upon and maintain shared knowledge in a non-trusted environment typical for blockchain. However, each piece of information is only shared within a subset of computing devices to which the information pertains. Yet, the term of ``blockchain'' is sometimes used to collectively refer to all systems in the domain including Corda.

We have been able to identify the following definitions in the literature:

\begin{asparadesc}
\item[Definition 1:] \emph{Blockchain is a system that uses the data structure of Bitcoin but extends the functionality}. 
This definition is used by, e.g., Bitcoin spin-offs that were created either due to hard forks or as an extension of the limited scripting functionality of Bitcoin. 
This definition is not limited, however, just to cryptocurrency systems; it can be utilized for a large spectrum of business logic by customizing the blockchain modules and protocols.
\item[Definition 2:] \emph{Blockchain is a system that maintains a chain of blocks}.
This definition allows for generalization of Definition 1: it allows data structures other than those used in Bitcoin. For example, Ethereum and Hyperledger match this definition.
\item[Definition 3:] \emph{Blockchain is a system that maintains a ledger of all transactions}.
The ledger does not need to be stored as a chain of blocks, however. IOTA is an example of a system that follows this definition.
\item[Definition 4:] \emph{Blockchain is a system with distributed non-trusting parties collaborating without a trusted intermediary}.
This definition rather refers to the main beneficial property of the paradigm. It was originally advocated by Corda~\cite{corda}.
\item[Definition 5:] \emph{Blockchain is a system that provides support for smart contracts}. Many blogs and popular science articles (such as~\cite{Chainlink-2023}) regard blockchain as a way of replacing paper-based contracts and human intermediaries with smart contracts, without considering how such contracts are implemented.
\end{asparadesc}


The first three definitions above are sorted by generality, from the most concrete to the most general. While they refer to the way the system is built, the last two definitions are about the way the system is used. 

It is important to distinguish between definition of blockchain and its characterization. While the definition has not been universally agreed upon, the fundamental properties have been extensively explored in the prior literature. For example, blockchain data is immutable: new data can be added but already included data cannot be deleted or modified. Blockchain additionally provides tamper-resilience, i.e., protection of blockchain data against any unwanted modification. Since immutability and tamper-resilience are explained and discussed in a large body of literature~\cite{gartner}, we do not cover them in this paper.


In the absence of proper definition, a blockchain is sometimes compared to a distributed tamper-resilient database with immutable data. It is important to observe that a blockchain differs from such a database in two fundamental ways. First, a database is an organized collection of data representing the current system state. The main functionality of a database is to allow efficient data retrieval, fusion, and aggregation triggered by user queries. In contrast, most blockchain implementations represent a ledger in which a history of transactions (or, more generally, of changes to the system state) is recorded. For example, there is simply no concept of a user balance in Bitcoin! While Ethereum keeps track of a contract state, it only provides limited means to retrieve and process state data, as explicitly defined by the contract. It does not support abstractions of a flexible query language, data view, schema, join, etc.
Besides, Ethereum records a history of all changes to the state, which results in blockchain space being more expensive and the storage less efficient compared to a database. 
As a result, only specific data elements (such as short transactions or indices) are stored on a blockchain. Many blockchain systems 
combine blockchain with offchain storage (databases or dedicated file systems). 

Secondly, the trust model is radically different as observed in~\cite{corda}. The database servers typically trust each other, even in federated databases, in the sense that they do not expect attacks from within the system. The main security focus is on making it difficult to compromise a server in the first place. To this end, database systems defend against malicious clients by using firewalls, strict access control, and many other methods. The situation is fundamentally different in the blockchain environment: the interests of participating computing devices are inherently misaligned so that they need to verify information received from each other and run a consensus to agree on changes to the data. While being able to agree on changes and progress in absence of a trusted administrator is a powerful abstraction, it bears a cost tag in terms of performance. If there are no misaligned interests between the participants and attacks from within the system are unlikely, there is little point in using blockchain technologies.


\begin{figure*}
	\centering
	\includegraphics[width=1\textwidth]{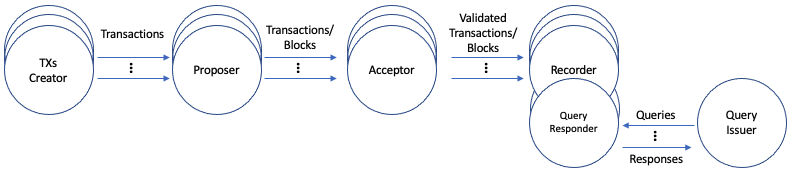}
	\caption{Roles of the participating computing devices and the relations between these roles}
	\label{fig:roles} 
\end{figure*}

\subsection{Blockchain types}


The two most important characterizations are the constraints on which participants are allowed to propose updates to the blockchain\footnote{Throughout this paper, by updating the blockchain we mean appending new data. Other types of updates are impossible due to data immutability.} and which participants are allowed to read blockchain data. In Bitcoin, any computing device may propose updates to the ledger or read it. This is also the case for the public deployments of Ethereum and IOTA. In Hyperledger Fabric, only authorized computing devices are allowed either. However, mixed models are also possible: In Ripple, every computing device has read access while only authorized computing devices can propose updates. We call these characterizations \emph{update-access-restricted} and \emph{query-access-restricted}, respectively. We refer to a system that is update-access-restricted or query-access-restricted as \emph{access-restricted}. Obviously, systems that are access-restricted require computing device authorization.

Usually, the identities are also handled differently in systems that are access-restricted. Namely, such systems typically use real computing device identities while systems that are not access-restricted commonly use public keys for identification. In principle, however, systems that are not access-restricted can still use real identities, if such identities can be verified and if the transparency of participation is more important than privacy. It is also not unimaginable for access-restricted systems to employ public keys or pseudonyms, though it may require the authorization or authentication component to operate under different trust and security assumptions compared to the blockchain system itself.

There exists another related dimension for classifying blockchain systems, namely \emph{decentralization}. Large-scale decentralized blockchain systems, such as Bitcoin, Ethereum, and IOTA, may have up to hundreds of thousands participating computing devices~\cite{Canellis2019,hindawi} without any coordinated management of the system structure, organization of computing devices, or network connections. On the other hand, consortium blockchain systems such as Hyperledger Fabric are smaller proprietary deployments where the consortium may decide to partially manage all of the above. We observe, however, that even consortium blockchain does not lend itself to complete management. In other words, all blockchain systems are self-organizing, though the extent of self-organization varies across the systems. The scale and extent of management significantly affects design priorities and implementation components as we explain in the rest of this survey. Consortium systems are access-restricted by nature. Large-scale decentralized systems are not query-access-restricted, though they may be update-access-restricted.

Unfortunately, the above complexity is not reflected in the currently existing terminology. 
All blockchain systems are coarsely divided into two categories in the existing literature. The first category is referred to as public or open or permissionless while the second category is called private or closed or permissioned. The exact meaning of these six terms and differences between them are not precisely defined to the best of our knowledge.


\subsection{Participants and their Roles}
\label{sec:participants}

Since the models and implementations significantly differ in various blockchain systems, we need to identify common fundamental elements in order to perform a systematic comparison. Two such unifying elements are the roles of participating computing devices and the conceptual layered architecture. We discuss these two elements in Sections~\ref{sec:participants} and~\ref{sec:architecture}, respectively. They apply to all blockchain systems studied in the rest of the paper, and they will likely generalize to many other blockchain approaches as well.


As commonly accepted in descriptions of distributed architectures, roles refer to the functional responsibilities. The same computing device may play a single or multiple roles in the system. The roles in a blockchain system are presented in Figure~\ref{fig:roles}.  


\begin{asparadesc}
\item[Creators of Transactions:] Different entities implementing a blockchain application can create transactions and inject them into the system by relaying them to the proposers.
\item[Proposers and Acceptors:] 
A central functionality of a blockchain system is to validate injected transactions and decide which transactions will be appended to the blockchain and in what order. This is the main responsibility of computing devices acting as acceptors. To this end, they need to run a distributed consensus protocol. However, all consensus solutions have an inherent limitation when it comes to scalability: they do not work very well if there are too many acceptors or too many concurrent transactions to be considered. This also makes consensus protocols susceptible to denial-of-service (DoS) attacks: an attacker can bombard a blockchain system with invalid transactions, effectively stalling consensus progress. To improve the scalability and resilience to DoS, most blockchain systems introduce the role of a proposer. Proposers act as intermediaries between creators of transactions and acceptors. They may reduce the rate of concurrent proposals by (a) verifying and pre-authorizing them locally and filtering out invalid or non-authorized transactions, (b) introducing explicit rate control, and (c) batching multiple transactions into a block. The exact distribution of responsibilities between the acceptor and proposer roles depends on a specific blockchain system but the conceptual separation applies to most systems. Since blockchain systems do not assume that all computing devices are trustworthy, they may need to incentivize the blockchain participants to perform their roles correctly, without deviations.
Additionally, proposers and acceptors contribute to making the blockchain tamperproof, together with data recorders. 

\item[Data recorders:] These entities record the additions to the blockchain accepted by the acceptors, which results in storing the entire blockchain. Along with  proposers and acceptors, they contribute to data security by making the data tamperproof through the use of cryptographic primitives.
\item[Query issuers:] They issue queries of different types over the current blockchain data.
\item[Query responders:] Some of the computing devices playing the role of data recorders have an additional responsibility of responding to the queries from query issuers based on the stored blockchain data. The need for and the exact role of query responders is further detailed in Section~\ref{sec:querying}.
\end{asparadesc} 

\subsection{Architecture}
\label{sec:architecture}

The design of blockchain systems is based on a layered architecture, which we show in Figure~\ref{fig:layers}.

\begin{asparadesc}
\item[Hardware Layer:] This is the bottommost layer of a blockchain system. While most blockchain systems can be deployed without any specialized hardware, hardware can make the computation (e.g., of cryptographic hashes) more efficient, provide extra security of the storage and computing environment, etc.

\item[Data Storage Layer:] This is the most important part of a blockchain system when it comes to storing data, keeping it safe from modifications, and making it traceable. This layer is also responsible for providing availability and durability of data. All of these features depend on the data items used in this layer, the method of storing the data, and the structure of the storage.

\begin{figure}[h]
	\centering
	\includegraphics[width=0.4\textwidth,scale=1]{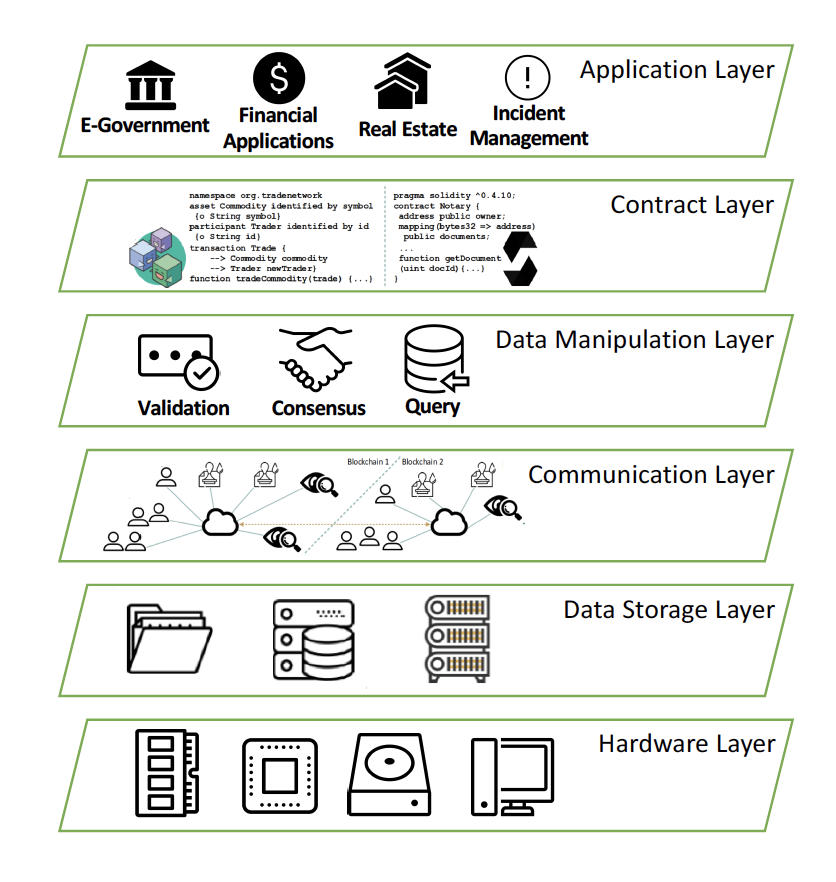}
	\caption{\label{fig:layers}Layers of a blockchain system}
\end{figure}

\item[Communication Layer:] 
%
Any blockchain system needs a mechanism for disseminating the transactions and blocks between the participants. This is especially important for large-scale permissionless blockchain systems. The granularity of dissemination, protocol of communication, ordering guarantees, privacy and security guarantees, and propagation time are related to this layer. All computing devices participating in a deployment of a blockchain system can be divided into two categories according to their involvement in the dissemination: core nodes constituting the blockchain network and ``client'' computing devices. Core nodes cooperate based on a common protocol to maintain blockchain data. In contrast, client computing devices do not play an active role in the dissemination. Instead, they get specific blockchain data of interest from core nodes.

\item[Data Manipulation Layer:] The main responsibilities of the data manipulation layer include updating the blockchain and offering a search functionality for blockchain data. Since an update must be a coordinated decision in a decentralized environment with a lack of trust between individual participants, the consensus protocol plays a central role in all blockchain systems. Consensus protocols vary across blockchain systems~\cite{Marko}, though the two most common categories are probabilistic Nakamoto consensus and BFT. Design choices of the data manipulation layer can affect different parameters related to performance and the strength of security and privacy guarantees, including the ability to withstand or mitigate DoS attacks.

\item[Contract Layer:] Blockchain systems provide the ability of defining contracts for creating and enforcing the rules among the participants of the network. Different blockchain systems may use various methods for developing the contracts based on their application domain. This layer deals with invoking and executing transactions within the contract, as well as programming languages (which may be Turing-complete or not) and execution environments.
    
\item[Application Layer:] This is an abstraction of the application built atop a blockchain system. While its specific design is outside the scope of a blockchain system, it defines the requirements that a blockchain system needs to satisfy.

\end{asparadesc}

\section{Understanding blockchain systems: overview and entities}
\label{sec:representative}

While pursuing the question of in-depth understanding of blockchain, we opt to illustrate the range of possibilities by analyzing four blockchain systems with significantly different properties and implementations. 
In this work, Bitcoin Core, Ethereum 1.0, Hyperledger Fabric v.2.0, and IOTA IRI v.1.5 are compared to each other in terms of blockchain entities and features and mechanisms of each architecture layer. The blockchain field is known for dynamic developments, with relatively short update cycles. However, Bitcoin does not undergo significant changes at this point so that we expect all Bitcoin-related descriptions to remain valid in the future.
Ethereum has declared its intent to release a new version with several significant changes, including a transition from PoW to PoS. However, no studies of the new version have been published ahead of the release so that we base our survey on the current operational version.
While Hyperledger Fabric continues to be under active development, the new versions introduce new features and performance optimizations, switch to a different default consensus implementation, and modify the chaincode features and structure. They do not radically change the fundamental design of each layer, which we discuss in our survey. 
Finally, IOTA IRI is currently the best documented version of IOTA that is presented in the white paper~\cite{tangle}.

\makeatletter
\providecommand*{\input@path}{}
\edef\input@path{{sections/}\input@path}
\makeatother


 Table~\ref{tab:roles} maps the blockchain roles to the entities of different blockchain implementations. In the following, we explain the functionality of each entity and the interactions happening between the entities.
 
   \begin{table*}
 	\setlength\tabcolsep{3pt} \setcellgapes{3pt} \makegapedcells
 	\centering
 	\caption{Mapping roles to the entities of different blockchain systems}
 	\label{tab:roles}
 	\scalebox{0.85}{
 		\begin{tabular}{|Y{2.3cm}|Y{2.4cm}|Y{3.5cm}|Y{2cm}|Y{2.5cm}|Y{3.5cm}|Y{2.2cm}|} 
 			\hline
 			\textbf{Blockchain system} & \textbf{Transaction creator} & \textbf{Proposer} & \textbf{Acceptor} & \textbf{Recorder} & \textbf{Query responder} & \textbf{Query issuer} \\
 			\hline
 			Bitcoin &  Nodes  & Miners &  Full nodes &  Full nodes & Dedicated services on top of full nodes& Everyone\footnotemark\\ 
 			\hline
 			Ethereum &  Nodes &  Miners & Full nodes & Full nodes & Dedicated services on top of full nodes & Everyone\\ 
 			\hline
 			Hyperledger Fabric&  Application clients &  Application clients, endorsers \& orderers & Orderers \& peers  & Peers & Application-dependent & Application clients\\
 			\hline
 			IOTA & Clients & Nodes & Nodes & Nodes \& permanodes & Dedicated services on top of nodes \& permanodes & Everyone\\ 
 			\hline
 	\end{tabular}}
 	\begin{flushleft}
 		\footnotesize{\textsuperscript{1} Everyone means any networked device}
 	\end{flushleft}
 \end{table*}
 
\subsection{Bitcoin}
\label{sec:bitcoin}
In the Bitcoin deployment, neither update access nor query access is restricted.
%
Bitcoin participants are called Bitcoin nodes in the Bitcoin terminology. They are split into two categories: full and lightweight. Some roles are only performed by full nodes, which act as both block acceptors and recorders of the whole blockchain. Upon receiving a newly broadcast block from the network, they verify it using the previously stored data and potentially append it to the blockchain. Additionally, full nodes bear higher responsibility w.r.t. dissemination in the Bitcoin network as we explain in Section~\ref{sec:communication}. A subset of full nodes is designated as miners that continuously attempt to create and propose new blocks.

Lightweight nodes are also called simplified payment verification (SPV)~\cite{nakamoto} nodes. SPV nodes neither store the entire blockchain nor participate in the decision of accepting new blocks. They can however, use the SPV method to verify that a block header is valid and that a given transaction is included into the blockchain. To this end, they can retrieve block headers and specific secure hashes from a full node and employ a Merkle verfication algorithm. Merkle verification is explained in Section~\ref{sec:datastorage}.

Both full and lightweight nodes are able to create transactions, thereby playing the role of a transaction creator. This effectively means that the separation between the entities is more blurred in Bitcoin (and similarly in Ethereum) compared to say, Hyperledger Fabric and IOTA. In particular, both terms of ``lightweight node'' and ``lightweight client'' are used in Bitcoin to refer to the same entities. 


Transactions are propagated to the miners that play the role of proposers.
As common for large-scale deployments, the Bitcoin network cannot sustain every miner making a proposal at an arbitrary point in time because of the scalability barrier and potential DoS attacks by the miners. It is therefore needed to moderate the number of proposers and rate of concurrent proposals. The entire Bitcoin system must attain a balance with respect to the proposal rate: a high rate will hamper scalability while a low rate will limit progress and transaction throughput. Furthermore, Bitcoin must achieve that moderation through distributed mechanisms, without a centralized moderating entity. In order to keep the rate down, Bitcoin introduces a \emph{Proof-of-work (PoW)} mechanism: When forming a block, a miner needs to solve a cryptopuzzle by performing a heavy computation and showing the proof of it, in order for the block to be accepted by other full nodes. 

For the Bitcoin system, there are dedicated services which act as query responders: they provide information for any computing device on the Internet even if the device is not on the Bitcoin network. However, these services can only handle fixed basic query types so that they are not as semantics-rich as specialized databases. Additional information about queries in blockchain systems is given in Section~\ref{sec:datamanipulation}.

\subsection{Ethereum}
\label{sec:ethereum}
The entities in the Ethereum system and the interactions between them are similar to those in Bitcoin. However, Ethereum introduces a number of additional underlying concepts and mechanisms not present in Bitcoin. The most central of those concepts is that of gas: since transactions in Ethereum can be Turing-complete programs, it is impossible to predict how much of computational and storage resources a transaction will consume. It is only feasible to monitor resource consumption at runtime and abort the transaction if it exceeds the permitted budget. Ethereum manages budgets for executable units such as transactions, blocks, smart contracts, etc. These budgets are measured in gas and paid for by the creators of transactions. Ethereum assigns gas cost to resources (operations and storage units). When a transaction executes an operation, Ethereum deducts the cost of that operation from the budget. 
In Section~\ref{sec:comparelayers}, we consider additional concepts and mechanisms in the context of each layer and cover them in detail.

\subsection{Hyperledger Fabric}
\label{sec:hyperledger}


As Hyperledger Fabric is designed for use in private blockchain systems, it does not have the concept of miners. Furthermore, its entities and their responsibilities are completely different compared to Bitcoin and Ethereum. 
There are three types of entities in Fabric: \emph{application clients, peers, and orderers}. Clients are computing devices that constitute the application running atop blockchain and play the role of transaction creators. They are external to Fabric and they connect to participants in the Fabric network. The peers are authenticated and authorized participants in the Fabric network that constitute the core of the private blockchain network. The orderers are dedicated computing devices provided by the Fabric deployment.

The flow in Fabric is as follows:
Clients send every proposed transaction to the \emph{endorsing peers} (also called \emph{endorsers}), which are a subset of all peers determined by the endorsement policy. More precisely, the endorsement policy defines the smallest set of endorsers that need to endorse a transaction in order for it to be valid~\cite{endorsement}. The endorsers typically come from different organizations within the consortium that deploys an application. 


Endorsers are responsible for simulating the transaction without updating the blockchain and verifying if the update causes a conflict with the current state or with the policy. If an endorser approves the transaction, it identifies the set of data read and updated by the transaction, which is called a read-write (or R-W) set. Then, the endorser sends the R-W set back to the client, along with the endorser's signature. If transaction approvals obtained by the client are insufficient according to the endorsement policy, the flow effectively terminates. Otherwise, the client submits the endorsed transaction and the R-W set to the \emph{ordering service}. 
%
%
%
%

Ordering service consists of orderers which are responsible for receiving the endorsed transactions and their R-W sets from different clients, ordering them, and forming a block. The orderers run a consensus protocol to achieve ordering of transactions.
Once produced by the consensus protocol, a block is disseminated to all of the peers in the system. Peers validate the block's transactions by checking whether the R-W sets still match the current state of the blockchain. Finally, the peers add successfully verified blocks to the blockchain and update the state. 

All peers store blocks, thereby playing the role of data recorders. There is no equivalent of Bitcoin and Ethereum lightweight nodes in Fabric. Orderers may be configured to store blocks as well.

In the above flow, the role of acceptors is distributed between endorsing peers in the first phase, orderers in the second phase, and all peers in the third phase. For example, the peers may reject the block produced by the orderers, if it is malformed or if it causes an inconsistency.

Since the deployment scale of private blockchains such as Fabric is typically smaller compared to a public blockchain such as Bitcoin, the significance of having a proposer role is reduced accordingly, see Section~\ref{sec:participants}. However measures towards improving scalability are still taken: Endorsers perform pre-authorization of transactions, clients themselves may abort the flow if the approvals are insufficient as per the endorsement policy, while orderers batch transactions and form blocks. 

Regarding query issuers and responders, application clients issue queries in Fabric, while the peers which keep the state respond to those queries. 

\subsection{IOTA}
\label{sec:iota}


Similar to Bitcoin and Ethereum, IOTA is a permissionless blockchain system. IOTA organizes transactions into bundles instead of blocks. The main difference is that a bundle groups related transactions together; a single transaction in a bundle cannot be understood or performed independently of other transactions in the bundle. For example, IOTA differentiates between input transactions that may combine funds from multiple input addresses together and output transactions that may split a sum into multiple output addresses. A bundle commonly contains related input and output transactions, as well as transactions of other types. More details are given in Section~\ref{sec:datastorage}.

The main purpose of grouping transactions in a block in Bitcoin, Ethereum, and Hyperledger Fabric is to improve the throughput of the consensus protocol that is used to update the ledger. Bundles cannot be used to this end because they cannot combine unrelated transactions. Interestingly, IOTA achieves higher throughput compared to Bitcoin and Ethereum by organizing bundles in a directed acyclic graph (DAG) as opposed to organizing blocks in a chain. We provide a detailed discussion in the context of the storage and data manipulation layers in Sections~\ref{sec:datastorage} and~\ref{sec:datamanipulation}.

The roles and responsibilities in the IOTA deployment are shared by three main entities: a) clients b) nodes and c) permanodes. Clients are external to the IOTA network. They connect to IOTA nodes via the HTTP API and interact with the network through the nodes. A client is responsible for creating transactions and forwarding them to a node.   


Nodes are interconnected together to form the core of the IOTA network. After having received a transaction from a client, a node needs to validate it. If the transaction passes the validation, the node will create a bundle as we discuss later, in Section~\ref{sec:manipulation-iota}. To prevent spamming and DoS attacks, nodes have to solve low-complexity cryptopuzzles in order to propose bundles. Having performed the low-complexity PoW, the node attaches the bundle to the local copy of the ledger called the Tangle~\cite{tangle} and propagates the bundle to the network. 
Nodes run a consensus protocol to synchronize their copies of the Tangle. 

Note that unlike Bitcoin and Ethereum, nodes do not create competing proposals for updating the ledger. Accordingly, there is no concept of mining in IOTA. Since nodes do not need to perform any significant computation, they are not rewarded for participation in the system either. Nodes are also able to answer limited queries about the Tangle but this is not their primary purpose.


Permanodes are dedicated nodes, many of which are provided by the IOTA Foundation itself. Similarly to nodes, they store a copy of the Tangle. However, they are external entities to the IOTA network: they do not receive bundles from the clients or participate in the consensus protocol. Instead, they receive Tangle updates from nodes. Their main purpose is to support complex queries. To this end they store Tangle information in a special database, as explained in detail in Section~\ref{sec:datastorage}.


In summary, clients play the role of transaction creators and proposers. 
Nodes play the role of acceptors (validating the transactions in the bundle), recorders (storing the bundles in the local ledger), and query responders (supporting services with simple queries). Permanodes play the role of recorders and query responders; yet they support richer queries compared to nodes.

\section{Understanding blockchain systems layer by layer}
\label{sec:comparelayers}

As we discussed in Section~\ref{sec:architecture}, blockchain systems are categorized into six different layers: hardware, data storage, communication, data manipulation, contract, and application. We now analyze the features and implementation of each layer in the four selected blockchain systems.

\begin{table*}
	\setlength\tabcolsep{3pt} 
	\makegapedcells
	\centering
	\caption{Hardware layer of different blockchain systems}
	\label{tab:hardware}
	\scalebox{0.85}{
      \begin{tabular}{|Y{4.5cm}|Y{3.6cm}|Y{3cm}|Y{3.1cm}|Y{4.7cm}| @{}}
		 \hline
		 \textbf{Features} & \textbf{Bitcoin} & \textbf{Ethereum} & \textbf{Hyperledger Fabric} & \textbf{IOTA} \\
		  \hline
		 Limiting resource & Processor & Memory bandwidth & Application-dependent & Processor \& network bandwidth\\ 
		 \hline
		 Cryptopuzzle solving device & ASIC & GPU & No cryptopuzzles 	& Proprietary processor (being phased out) \\ 
		 \hline
Additional hardware for security (Research initiatives) & Hardware-based trusted execution environment (e.g. Intel SGX processor) & Hardware-based trusted execution environment & Application-dependent &  No additional hardware \\
		 \hline
		\end{tabular}}
\end{table*}

\subsection{Hardware Layer}
\label{sec:hardware}

While blockchain implementations are primarily software-based, hardware components are used for two purposes: improved efficiency and enhanced security. In particular, implementations of open blockchain systems such as Bitcoin, Ethereum, and IOTA, are resource-bound: there is a hardware resource that limits the ability of a single proposer to propose numerous changes to the blockchain at a fast rate. The exact resource varies across the systems, however. On the other hand, this is not required in permissioned blockchain because of a tight membership control.
Table~\ref{tab:hardware} highlights related aspects for the analyzed systems. 

As mentioned in Section~\ref{sec:representative}, miners in Bitcoin require to perform costly and time consuming computations in order to create a valid block. They need to solve a cryptopuzzle by exhaustively going over the solution space and calculating a hash value for each potential solution, which results in a massive amount of hash computations.  
Therefore, the Bitcoin technology is strongly dependent on the processing power of the miners.
Nowadays, the most common hardware for the purpose of mining bitcoins are ASIC processors~\cite{bitcoinText}. They are designed specifically for the purpose of computing Bitcoin hashes in an optimized way.

In contrast to Bitcoin, Ethereum follows ASIC-resistant approach in its hardware layer. The goal of choosing such an approach is coming up with a puzzle that reduces the gap between the most cost-effective customized hardware and what most general-purpose computers can do, so that it would be economical for individual users to mine with the computers they already have~\cite{bitcoinText}. To achieve this goal, 
Ethereum uses a different PoW algorithm compared to Bitcoin, called ethash~\cite{ethash}. 
The ethash algorithm needs 64 sequential page fetches from the memory to generate a single hash and compare it with the cryptopuzzle target. Since ethash is bound by the speed of memory access rather than computation, speeding up the processor computation by ASICs does not help in a significant way. Furthermore, since an expensive top performing computer only has moderate improvement in the speed of memory access compared to commodity hardware, vertical scaling does not have a strong effect on the efficiency of mining in Ethereum. 
Hence, utilizing GPU processors, which can solve cryptopuzzles faster than CPUs~\cite{gpu}, is more cost-effective for mining in Ethereum. This has an additional effect that 
cryptopuzzles in Ethereum can be less computationally intensive. On the other hand, the power consumption of GPU is higher compared to ASIC so that the net effect on power consumption in Ethereum is not clear.

The design of cryptopuzzles in both Bitcoin and Ethereum facilitates parallel computation, which increases the risk of mining power centralization in the hands of powerful players. While ethash limits the potential of vertical scaling for mining, horizontal scaling is widely used in both systems.

As there is no concept of mining in Hyperledger Fabric, it does not need specialized hardware components to boost the efficiency, unless required by a specific application. 
While IOTA does not use blocks and does not employ the concept of mining or monetary rewards, it still uses a less computationally intensive version of cryptopuzzles in order to prevent denial-of-service attacks. Therefore, IOTA is also dependent on the computation power of participating nodes. Since resource-limited IoT devices is the main focus of IOTA implementations~\cite{iot}, IOTA has developed a proprietary low-energy processor called JINN~\cite{jinn}. The main purpose is to expedite the computation of relevant cryptographic primitives, though currently the computation is typically done in software.
Besides computation power, bandwidth is another critical resource in IOTA since IoT devices are also bandwidth-limited~\cite{networkBound}.

A major focus of the blockchain technology is to provide tamperproof storage in absence of trust in individual participants.
In view of this,
a number of initiatives such as Teechain~\cite{teechain} have tried to enhance the security of Bitcoin and Ethereum by taking advantage of the hardware-based Trusted Execution Environment (TEE) technology, such as Intel\textsuperscript{\textregistered} SGX processor~\cite{intel}.
TEE is designed to create a more secure computation environment for the processor by isolating and protecting the running application against unauthorized access or tamper by the host machine. While the use of TEE still requires the trust in the TEE manufacturer, it mitigates potential attacks by individual blockchain participants.


\subsection{Data Storage Layer}
\label{sec:datastorage}

%
%

The purpose of the storage layer is to record all transactions in a distributed ledger and provide support for their efficient verification. While simple transactions merely transfer financial tokens, general transactions represent transitions in the global system state. It is important to keep track of the state, e.g., in order to perform verification of transactions. While it is possible to reconstruct the most updated state by starting from the initial state and replaying all transactions recorded in the ledger, this would be a time-consuming and inefficient process, especially since the ledger size is continuously growing. As a result, all blockchain systems store explicit information about the current state in addition to the transactions. However, the systems differ in terms of what state-related information they store and in terms of how they organize it.
Additionally, the storage layer provides support for verification of transactions issued by the clients and for computation on the state in the ledger.

The rest of the storage layer description covers state tracking approach, general organization of blockchain storage, the block structure, structure of transactions and their grouping in blocks, on-disk storage, the use of trees in Bitcoin and Ethereum, in-memory storage, and data retention. Table~\ref{tab:storage} presents a comparison of salient storage design aspects across the four systems.

\subsubsection{State Tracking Approach}
\label{sec:state-tracking}



\begin{table*}
	\setlength\tabcolsep{3pt} \setcellgapes{3pt} \makegapedcells
	\centering
	\caption{Data storage layer of different blockchain systems}
	\label{tab:storage}
	\scalebox{1}{
		\begin{tabular}{|Y{1.2cm}|Y{1.3cm}|Y{3cm}|Y{3.2cm}|Y{3.4cm}|Y{3cm}|@{}}	
				\hline	
				\multicolumn{2}{|Y{2.5cm}|}{\textbf{Features}}& \textbf{Bitcoin} & \textbf{Ethereum} & \textbf{Hyperledger Fabric}  & \textbf{IOTA} \\
				\hline	
				\multicolumn{2}{|Y{2.5cm}|}{State tracking approach} & UTXO & Account-based & Application-dependent  & UTXO \\			 
				\hline
				\multicolumn{2}{|Y{2.5cm}|}{Higher-level structure} & Sequence of blocks with transactions & Sequence of blocks with transactions & Sequence of blocks with transactions   & DAG of transaction bundles with edges signifying approvals  \\			 
				\hline
				\multicolumn{2}{|Y{2.5cm}|}{Maximum block/bundle size} & 1 MB block fixed by the protocol  & Block's gas limit determined by miners  & Configurable block size as per application  & Unlimited bundle size \\			 
				\hline
				\multirow{4}{*}{\makecell{On-disk \\ storage}} & Ledger storage & Block data as multiple files of  limited size each & Block data as multiple files of limited size each  & Block data as multiple files of configured size  & Tangle stored in a RocksDB database \\
				\cdashline{2-6}
				& Index & Block index in a LevelDB database & - & Block index in a \newline LevelDB database & - \\
				\cdashline{2-6}
				& State-related storage & UTXO set in a LevelDB database  & Each vertex of tries (state, storage, receipts, txs) in a LevelDB database & Worldstate in a LevelDB or CouchDB database & Balance info stored in RocksDB database  \\
				\cdashline{2-6}
				& Extra storage elements & Each block file has a corresponding undo file to support reorganization/fork  & All block data (block header and transaction data) in a LevelDB database & - & Snapshot data as a file \\ 	  				 			 
				\hline
				\multicolumn{2}{|Y{2.5cm}|}{In-memory storage} & Block index database, UTXO cache, \& Merkle trees of transactions & Cache of Merkle Patricia tries vertices & Application-dependent; peer cache & Tangle Accelerator (proxy cache for IOTA nodes) \\			 
				\hline	
				\multicolumn{2}{|Y{2.5cm}|}{Data retention} & No retention policy & State tree pruning & No retention for ledger; application-dependent for the state  & Snapshot purges the database \\
				\hline
		\end{tabular}}
	\end{table*}

So far, all blockchain systems have been employing one of the two principal approaches for tracking system state. Bitcoin employs the unspent transactional output (UTXO) approach while Ethereum falls under  the arbitrary state approach. IOTA follows the UTXO approach with a minor modification, as we explain below. Fabric does not track state by default but the application developer can devise and plug in any desired implementation with some restrictions. The two approaches differ by generality, compactness of transactions, simplicity and efficiency of storage organization, ease of parallelizing transaction processing, as well as transaction linkability, i.e. that ease at which a transaction can be linked to an individual user.

In the UTXO approach, a transaction transfers currency tokens by consuming a number of input tokens and producing a number of output tokens. For example, suppose Alice has previously received 6 tokens in a single transaction. In this case, the six tokens become identifiable transactional outcome (TXO), which is marked as unspent. Assume Alice later wants to transfer 4 tokens to Bob. 
She can use the six token UTXO as the input to the new transaction.
During the transaction execution, the  protocol will verify that the UTXO has not been spent. Then, three new UTXOs will be created: one UTXO with transaction fee will be transfered to the block creator, one UTXO worth 4 tokens will be transferred to Bob while another UTXO worth two tokens minus transaction fee will be transferred back to Alice. 
Additionally, the old six token TXO will be marked as spent. 
In the UTXO approach, the system only keeps track of the unspent transactional outputs because they can be used as input to future transactions. 
Since there is no concept of accounts or wallets at the protocol level, the ``burden'' of maintaining a user's balance is shifted to the client side. Wallets maintain a record of all UTXOs associated with a user and compute the total sum, which represents the balance. However, this balance is inaccessible to all but the user who owns the wallet.

The UTXO approach in IOTA slightly differs in that the term of ``address'' is used to refer to a collection of multiple TXOs because an address can be reused to receive funds in multiple transactions, thus acting as a pseudo-account. The storage system in IOTA keeps track of all transactions that have transferred funds to a given address. Like a TXO, an address can only be used once when sending funds such that the entire amount is spent. This minor difference in IOTA does not have a significant impact on the properties of the UTXO approach, which we discuss below.



The arbitrary state approach is more general in that it supports arbitrary state rather than just tokens transferred between the users. This naturally leads to more complex storage structures compared to the UTXO approach. In particular, it is common to keep the state partitioned. For example, each user in Ethereum has an associated account and the state is partitioned by those accounts so that the state tracking approach in Ethereum is commonly referred to as ``account-based''. Since there exists no single public deployment of Hyperledger, the state in Hyperledger is kept separately for each proprietary deployment. Besides, the worldstate in Fabric is further partitioned per channel. The configurable nature of Hyperledger allows an application developer to choose an appropriate approach per partition, based on the application needs.

Since the general state may be of an arbitrarily large size, a question of limiting the state size arises. This challenge is mitigated in permissioned blockchains by the ability to have a tighter control over the behavior of each particular user and application. In public blockchains, on the other hand, the users need to be disincentivized from storing too large of a state. This is achieved in Ethereum through the gas spending mechanisms mentioned in Section~\ref{sec:ethereum}: the users pay gas not only for transaction execution but also for state storage, proportionally to the state size.

The per-account state in Ethereum  includes the user balance as well as any additional variables required for general calculations. In the above example, a transaction transferring four tokens from Alice to Bob would need to check Alice's balance, decrease it, and add the funds to Bob's balance. 

A single user may receive many token transfers in the UTXO approach and may own many unspent TXOs simultaneously. This means that when limited to applications that transfer currency tokens, the blockchain state may grow with the number of transactions over time, whereas in the account-based approach, the state size only depends on the number of users.  In this scenario, the account-based approach may require less storage compared to the UTXO approach. Besides, a single UTXO-based transaction may take many input TXOs and may produce many output TXOs. This makes the verification protocol less efficient compared to the simple verification procedure in the account-based approach.




On the other hand, the UTXO approach may allow for better parallelization of transaction processing. For example, if Alice wants to transfer one UTXO to Bob and another to Carol, the two transfers can be handled in parallel. In contrast, Alice's balance would need to be checked sequentially in the account-based model, which would require serialization of the two transfers.


The UTXO approach also has an edge when it comes to privacy, specifically hiding the link between the user and her transactions and balance. In the account-based approach, the system maintains the user's balance and keeps explicit association between the transactions and accounts. No such association is maintained in the UTXO approach so that the state cannot be linked to an individual user.

As mentioned in the beginning of this section, Fabric does not provide any  implementation for state tracking that would be included in the installation. The application developer, however, can implement any state tracking approach in Fabric using the key-value store provided to this end. If a financial application needs to maintain user accounts with corresponding balances, the account name can be used as a key while the balance will be included in the value. The UTXO model can be implemented in the key-value store of Fabric as follows~\cite{fabric}. First, every UTXO can be represented as a unique key-value entry in the store. Any unique identifier created for the UTXO can be used as a key. The value will specify (a) the amount of cryptocurrency that the UTXO holds and (b) the reference to the owner of the UTXO, which can be represented in different ways, such as the public key or the Fabric identity. Any transfer transaction will spend old UTXOs and destroy their entries in the store. It will also create new entries, one for each new UTXO.

\subsubsection{Organization of blockchain storage}
\label{sec:storage-org}

The storage organization is conceptually similar in Bitcoin, Ethereum, and Fabric. The transactions are grouped together into blocks to improve scalability of the consensus protocol as explained in Section~\ref{sec:datamanipulation}. The blocks are stored in files on disk. To provide tamper resistance, the individual blocks are linked to each other to form the logical structure of a chain. Since each block includes a hash pointer to the previous block in the chain, tampering with one block would affect the hash values in all subsequent blocks in the chain. 

In addition to the chain of blocks, the storage includes a key-value store for keeping auxiliary information, e.g., related to the system state as explained in Section~\ref{sec:state-tracking}. An increased emphasis on the state storage and state-related operations (such as search) in Ethereum results in significantly more complex storage mechanisms compared to Bitcoin. Besides, the balance of roles between the chain of blocks and the key-value store is fundamentally different in Bitcoin and Ethereum because a key-value store lends itself better to state-related operations. In Bitcoin, search for blocks and even transactions is conducted primarily using the chain of blocks, with the help of a block index maintained in the key-value store. In Ethereum, on the other hand, block headers and transactions are duplicated in the key-value store, which is the main storage element used in the search. Furthermore, much of the data kept in the key-value store is organized in cryptographic search trees (henceforward referred to as cryptographic tries), while Bitcoin only constructs trees in memory and for a different purpose, as we elaborate upon in Section~\ref{sec:tries}.



IOTA is different from Bitcoin, Ethereum, and Fabric in three respects: there are no blocks, the structure is different from the linked list, and the transactions themselves are stored in a database rather than a ledger. As opposed to the chain of blocks, IOTA utilizes a directed acyclic graph (DAG). In essence, the vertices represent transactions while the edges signify approval of the transactions. A new transaction has to verify and approve two existing transactions in order to be included in the DAG. 
Accordingly, the consensus algorithm significantly differs as we explain in Section~\ref{sec:manipulation-iota}.
When it comes to storage, DAG storage requires more space compared to systems that use a chain of blocks. This is because there are more vertices in the DAG than blocks in a chain, and every vertex has non-trivial meta-information attached to it.
 
 


\subsubsection{The block structure in Bitcoin, Ethereum, and Fabric}
\label{sec:block-structure}

The key differences in the block storage structures in Bitcoin, Ethereum, and Fabric are mainly due to two important design choices: a) record keeping model and b) consensus protocols. As the account model requires more storage to represent the association between the accounts and the balances, the block structures for account models are relatively complex compared to those for UTXO. As we detail in Section~\ref{sec:on-disk}, Ethereum has to keep track of the states and changes associated with each account in contrast to UTXO-based Bitcoin.

The block body in all three systems includes a list of transactions. In Ethereum, it additionally includes a list of uncles (whose concept is explained in Section~\ref{sec:manipulation-ethereum}).

The block header in all three systems contains the parent block hash. This hash is inherent to all blockchain systems that create a chain of blocks and presents the cornerstone for the immutability property. The hash is only calculated after the block formation has been completed; thus, a hash of the block cannot be included in the block itself. Instead, Bitcoin, Ethereum, and Fabric nodes compute the block hash when they receive the block from the network. Fabric nodes, however, additionally store the hash of the block data in the block header.
      
Nonce is another important field in the block header of blockchain systems that use mining such as Bitcoin, Ethereum, and IOTA; it contains a solution for the block cryptopuzzle.
In Ethereum, the header additionally includes a hash of the uncle list stored in the block body, as explained above. 
Furthermore, Ethereum stores the root hash for each of the multiple search tries in the block header, as we explain in Section~\ref{sec:tries}.


%
%


Considering the permissioned nature of Hyperledger, the block storage structures in the system include the real identity and signatures of the clients, endorsing peers, and the orderer. Such identities are not present in Bitcoin, Ethereum, and IOTA. Furthermore, invalid transactions may be included as part of the confirmed block due to the working principles of the consensus protocol of the Hyperledger as explained in Section~\ref{sec:datamanipulation}.  Therefore, there is a filter flag for each stored transaction, which is used to differentiate between valid and invalid transactions in the block.


\subsubsection{Structure of transactions and their grouping in a block}
\label{sec:trans-struct}

The maximum number of transactions that can be grouped together is determined by the maximum size of the block.  For example, in Bitcoin, the maximum size of the block is fixed by the protocol at 1MB. The transaction size in Bitcoin varies since a UTXO transaction can have multiple inputs and multiple outputs.  Unlike Bitcoin,  Ethereum uses the concept of gas to determine the size of the block. Every transaction in the block has a gas price and the sum of prices for all transactions in a block should not exceed the maximum gas limit of the block set by Ethereum miners.  In the case of Hyperledger, the maximum block size can be configured by the administrator of the network,  in accordance with the application requirements.

 
While not using blocks, IOTA utilizes the concept of bundles. A bundle is a group of transactions that are tied to each other. The need for bundles arises because, unlike Bitcoin, transactions in IOTA are limited in terms of the maximum allowed number of inputs and outputs. Since proposing each transaction in a bundle requires a separate proof-of-work puzzle (see Section~\ref{sec:dos}), these limits translate to the maximum number of inputs and outputs permitted for each puzzle; longer bundles would thus require more work. Thus, if Alice wants to consolidate funds from multiple addresses and transfer those funds to several different users, she would need to create a bundle with multiple transactions.
Thus, the entire bundle is issued by the same client. The size of the bundle is determined by the input; it is unpredictable and beyond the system control. Furthermore, IOTA does not have a predefined limit on the number of transactions in the bundle. 

In the IOTA implementation presented in this survey~\cite{iri}, the bundles can be used as a vertex in the DAG as opposed to individual transactions.

\begin{figure}
	\includegraphics[width=85mm,scale=1]{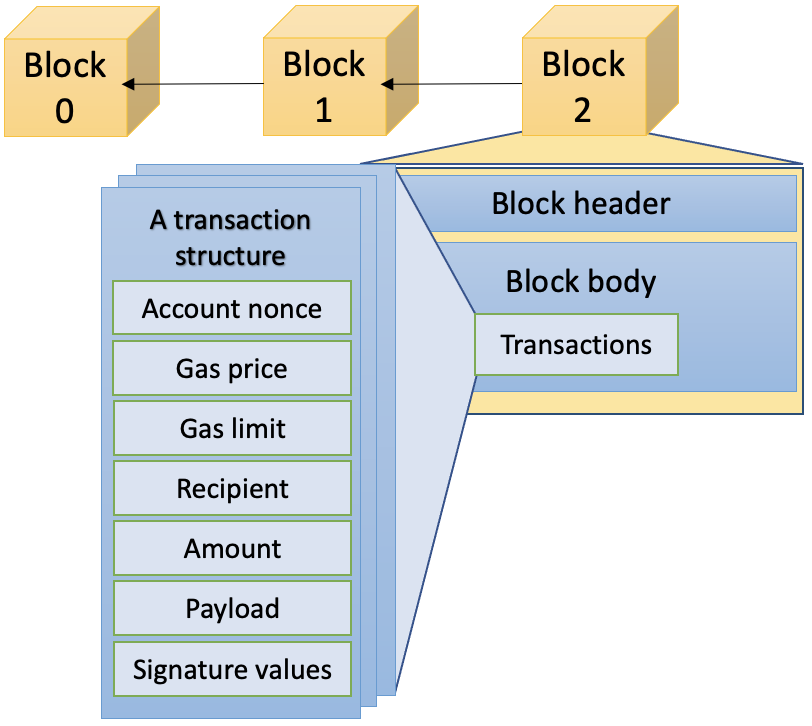}
	\caption{An Ethereum transaction structure}
	\label{fig:organization}
\end{figure}

Bitcoin transaction structure is described both in the specifications~\cite{bitcoinTransaction} and textbooks~\cite{bitcoinText} in detail.  Its main fields are the list of inputs and the list of outputs common in the UTXO model as explained in Section~\ref{sec:state-tracking}. As Ethereum does not follow the UTXO model, it has a different transaction structure. As shown in Figure~\ref{fig:organization}, the transaction structure in Ethereum includes the following information:

\begin{asparadesc}
\item[Account nonce] (called \emph{nonce} in Ethereum code~\cite{ethereum-tx-code}): An Ethereum node issuing transactions on behalf of the user maintains a counter of the number of transactions sent from the user's account. When the transaction is issued, the node sets account nonce to the current value of the counter. Keeping track of this count over time is important for two reasons: first, it establishes sequential order between transactions sending funds from the same account so that the miners or validators will process the transactions in that order. Second, the account nonce prevents transaction replay attacks by the recipient of funds because a repeated transaction with the same nonce will be detected as invalid.
\item[Gas price] (called \emph{gas\_price} in Ethereum code) is the price of each unit of gas, expressed in Ether. 
\item[Gas limit] (called \emph{gas} in Ethereum code) is the maximum amount of gas that can be consumed by executing this transaction. 
\item[Recipient] (called \emph{to} in Ethereum code) is the address of the transfer receiver. 
\item[Amount] (called \emph{value} in Ethereum code) is the value to be transferred to the recipient. 
\item[Payload] (called \emph{data} in Ethereum code): If the transaction is meant to be an execution of a contract (the recipient is a contract account), the payload field would be a message that identifies the function and argument values of the contract. Otherwise, if the transaction is for payment (the recipient is a user account), the payload field would be empty. 
\item[Signature values:] This field includes components of the signature algorithm of the sender. Ethereum transactions use ECDSA~(Elliptic Curve Digital Signature Algorithm)~\cite{ecdsa} as their digital signature for verification. The signature of a transaction confirms that the sender has authorized this transaction. 
\end{asparadesc}

In Hyperledger Fabric, the design of the transaction structure depends on the application and state tracking approach chosen for the specific deployment.

\subsubsection{On-disk storage}
\label{sec:on-disk}

The most important element stored on the disk is the ledger. Bitcoin, Ethereum, and Hyperledger Fabric store their block data as multiple files, each of which is limited in size. Although the file size limit is fixed in Bitcoin and Ethereum, it is configurable in Hyperledger Fabric~\cite{fabricBlockfile}. Block files store the blocks appended to the local copy of the ledger~\cite{bitcoinStorage}. 

 \begin{figure*}
	\includegraphics[width=175mm,scale=1]{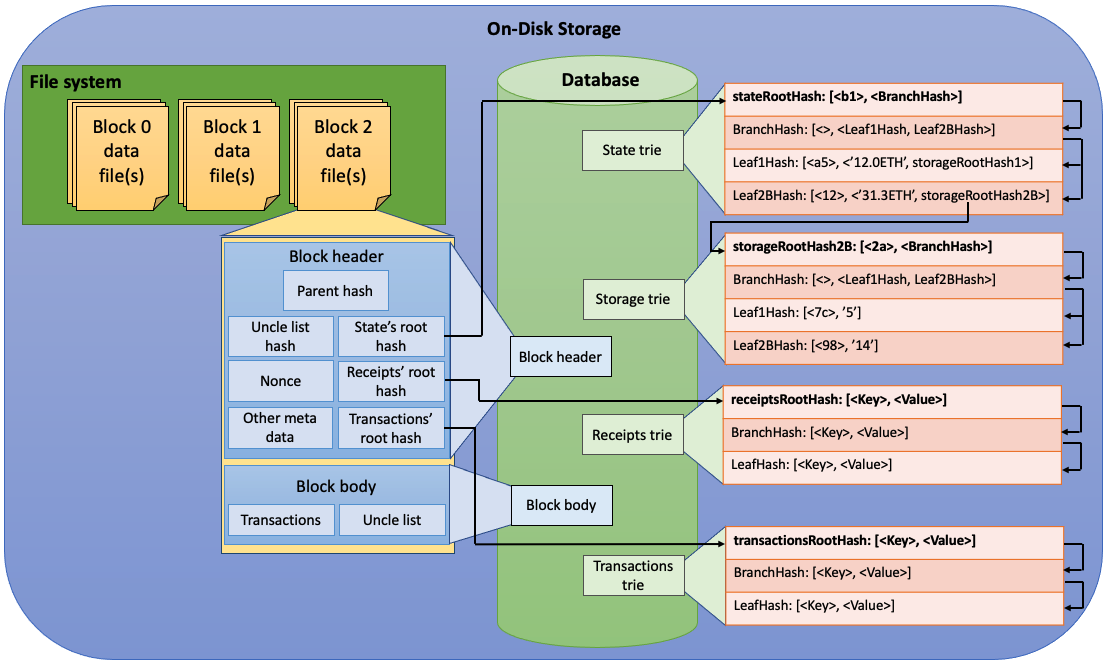}
	\centering
	\caption{On-disk storage of Ethereum consists of the block data and database records}
	\label{fig:disk}
\end{figure*}  

Bitcoin additionally stores block indexes on the disk to keep track of the information available in the chain of blocks. These indexes are stored in a LevelDB (key-value store) database. Similar to Bitcoin, Hyperledger Fabric utilizes indexes. By default, Fabric also stores indexes in LevelDB but it can be replaced by other databases. The primary purpose of the indexes is to support search functionalities for blocks and transactions stored in the file through various keys such as a block number, block hash, transaction id, and transaction number. The values of the LevelDB key-value store are corresponding file location pointers to the chain of block files~\cite{bitcoinStorage,fabriccode}. When an API call to access blocks and transactions of the blockchain occurs, these search indexes are utilized to find the actual data location in the file.
On the other hand, Ethereum relies on a database for keeping track of changes and finding data items, so that has no need for indexing the files.

State-related storage is another salient on-disk element. Bitcoin, Ethereum, and Hyperledger Fabric use a database of key-value pairs for keeping track of the latest blockchain state.
In Bitcoin, the UTXO set which stores all unspent outputs is all information that is required to validate a new transaction without the need to traverse the whole blockchain~\cite{delgado2018analysis}. Bitcoin keeps the UTXO set in a LevelDB storage called \emph{chainstate}. On the other hand, Hyperledger Fabric keeps track of the latest values of the business assets in its state-related storage called \emph{worldstate}. 
The state-related storage of Hyperledger Fabric is represented on the disk as a LevelDB key-value store by default. The keys and values of the state-related storage reflect the data model described in the application \emph{chaincode}, as defined by the developer (see Section~\ref{sec:contract}).
In Fabric, if the business application requires a complex data model and access pattern, various other databases such as CouchDB, GraphDB, and a relational datastore can be used to represent the worldstate instead of LevelDB. These alternatives support rich queries and data types~\cite{couchdb}.
Similarly, Ethereum stores the latest values of all the accounts in a LevelDB database as the state-related storage to facilitate calculating the current value of any account available in the ledger. For this purpose, Ethereum constructs the following four tries:
 
\begin{asparadesc}
	\item [State] trie represents the global state that efficiently stores the mapping between all the addresses and accounts. 
	\item [Storage] trie is responsible for maintaining the relationship between the account and the corresponding balance. State trie is linked to the storage trie. 
	\item [Transactions] trie represents the transactions that change the state of the Ethereum. 
	\item [Receipts] trie represents the outcome of the successful execution of transactions.  
\end{asparadesc}

Individual systems additionally maintain auxiliary meta-data stored on the disk.
In Bitcoin, each block file has a corresponding undo file which contains the information that is necessary to roll back and remove a block from the blockchain in the event of a reorganization/fork~\cite{bitcoinStorage}.  
In Ethereum, in addition to storing block data as files, all block data including the block header and the block body is stored in a LevelDB database as shown in Figure~\ref{fig:disk}. 


Due to the different organization of blockchain storage in IOTA as described in Section~\ref{sec:storage-org}, Tangle is stored in a database rather than in files. Thus, all the information about both transactions and balances is kept in the RocksDB database. As an extra element, an IOTA node keeps snapshot data (explained in Section~\ref{sec:retention}) in a file.
Each time a new snapshot arrives, it removes all transactions from the Tangle, i.e. RocksDB database, and moves only non-zero balances to a file in order to be used by the new pruned Tangle. 
Thus, similarly to Hyperledger Fabric, IOTA maintains only the current balances associated with the addresses.

\subsubsection{On the use of trees in Bitcoin and Ethereum}
\label{sec:tries}

Bitcoin and Ethereum use trees to organize data items (transactions and in the case of Ethereum, other data elements related to state tracking) in the context of a given block. The trees are cryptographic: every non-leaf vertex stores a secure hash value for each of its child vertices. This way, tampering with a single tree vertex $n$, or inserting a new vertex $n$ will lead to modification of all the hashes on the path from $n$ to the root of the tree, making it easily detectable. Due to this, the hash of the root vertex becomes a cryptographic fingerprint of the entire tree structure. Because of its significance, the hash of every root vertex is kept in the corresponding block of Bitcoin or Ethereum for the sake of verification.  

\begin{figure}
	\includegraphics[width=85mm,scale=1]{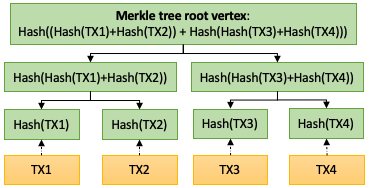}
	\centering
	\caption{Bitcoin's transaction Merkle tree}
	\label{fig:merkle}
\end{figure}

However, Bitcoin and Ethereum use trees for different purposes, and accordingly, their implementation is quite different. Bitcoin uses binary Merkle trees as shown in Figure~\ref{fig:merkle}. For a given block, the tree is constructed as follows: each leaf represents a block transaction. The order of leaves corresponds to the deterministic order in which the transactions are listed in the block body. 
Each non-leaf vertex is created by computing a secure hash of the concatenation of the two child values.
Due to this construction, the tree is deterministically reproducible from the block body stored on the disk. This is important because Bitcoin constructs Merkle trees in memory on demand.
 
The main purpose of utilizing Merkle trees in Bitcoin is to minimize the amount of bandwidth used for disseminating the proof that a given transaction is included in a given block. It is a typical situation that a Bitcoin lightweight node, without the knowledge of the ledger, inquires about the inclusion status of a particular transaction in a new block. To this end, the node sends a verification request including the transaction hash to a full node which keeps track of the ledger. The full node first needs to locate the Merkle tree of the block in memory and then, sequentially scan the leaves of the tree in order to verify the inclusion. While a sequential scan is not an efficient operation, its inefficiency does not have a strong impact because the number of transactions in a block is relatively small. However, once the transaction is found, the full node can construct the Merkle proof efficiently~\cite{merkle-proof}. The Merkle proof includes all vertices on the path from the leaf with the transaction to the root of the tree, along with the children of these vertices. The full node can then send the Merkle proof back to the lightweight node. The crucial point is that the lightweight node can verify the inclusion based on the Merkle proof, without requiring the rest of the vertices in the Merkle tree. The size of the Merkle proof is logarithmic with the size of the Merkle tree.

In addition to constructing Merkle proofs, there are in fact very few operations supported by Bitcoin Merkle trees. Miners benefit from being able to efficiently append new leaves with transactions because new transactions can arrive while mining a block. Merkle trees allow for efficiently implementing this particular type of vertex insertion because a new transaction becomes the rightmost leaf. Additionally, the trees can be pruned: for example, if we do not need to verify the inclusion of Transactions 1 and 2 in Figure~\ref{fig:merkle} any longer because their TXOs have already been spent, we can remove corresponding leaves, along with their hash vertices. We do need to keep the vertex with \textit{Hash(Hash(TX1)+Hash(TX2))}, however, for the purpose of constructing Merkle proofs, as explained above. On the other hand, vertex search, update, and deletion are neither needed nor efficiently supported.

This is in a stark contrast with Ethereum, which uses the trees for storing the state, in addition to transactions. Unlike transaction history which is fixed, the state in Ethereum needs to be frequently changed: new accounts are inserted and old accounts are deleted. The state of a particular account needs to be searched and updated as well, potentially often. Ethereum uses Merkle Patricia trie, which provides logarithmic complexity for search, insertion, update, and deletion, in addition to efficient construction of Merkle proofs. Similarly to the Merkle tree, data items are stored in the leaves. However, Merkle Patricia trie is a prefix tree that uses hashing mechanisms to produce a key string that determines the path of the data item within the tree.
Normally, prefix trees create a new tree level for each character in the key: the first character will determine the branch at the root, the second character will determine the branch at the next level, and so on. However, Merkle Patricia trie is a compact radix tree: if there are two vertices with keys \textit{``abcd''} and \textit{``abef''} respectively, an intermediate vertex with the key \textit{``ab''} will be created. This vertex will have two direct children corresponding to \textit{``abcd''} and \textit{``abef''}, without any additional intermediate levels. In Ethereum, the key is encoded as a hexadecimal value so that each vertex has at most 16 children.

The key string that determines the path is produced differently for each trie type.  
For the state trie, each leaf corresponds to an account in Ethereum. The key string is derived from the hash of the account address, while the value of the leaf includes the balance and the root hash for the storage trie of the account. 
For the storage trie, each leaf represents an Ethereum contract's data. The key string is derived from the hash of the Ethereum contract's address, while the values reflect the data model described in the Ethereum smart contract. An updated state of an Ethereum account is retrievable by traversing the storage trie vertices in the LevelDB database.



\begin{figure*}
	\includegraphics[width=175mm,scale=1]{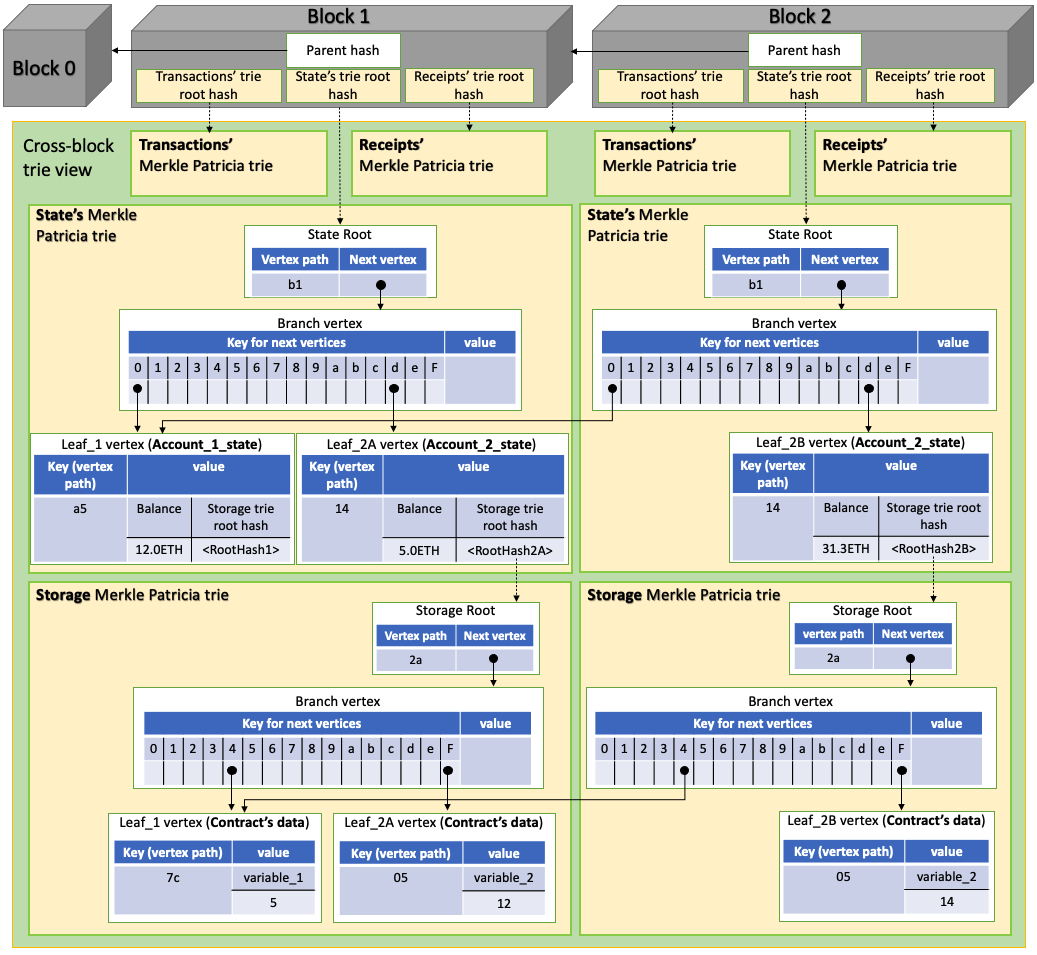}
	\centering
	\caption{Constructing tries across multiple blocks in Ethereum}
	\label{fig:memory}
\vspace{-4mm}
\end{figure*}  


As shown in Figure~\ref{fig:disk}, each vertex of every Ethereum Merkle Patricia trie is stored as a separate record in the LevelDB storage. The key for the record is derived from the hash of the trie vertex and stored in the parent vertex record. Since the hash for the root vertex is stored in the block header, it is used as the starting point for finding other trie vertices in the database, which allows for an efficient traversal of the entire trie.



Figure~\ref{fig:disk} provides an illustration for the storage organization on disk in Ethereum. Header and body for \emph{Block 2} are stored in both \emph{Block 2} data file(s) and in the database. The header contains root hashes for the state, receipt, and transaction tries. Each root hash is a key for the corresponding root vertex record in the database. From the root record of the state trie, we can reach the branch vertex record, which points to \emph{Leaves 1} and \emph{2B}, each of which corresponding to a different account. The record of \emph{Account 2B} points to the root of the storage trie for that account. The pointers between the vertices of the storage trie in the database are presented in a similar way.

Another interesting functionality of Ethereum is that it continues maintaining correct tries for old blocks, not just for the most recent block. For example, we can query the value of an account or the value of a variable (maintained by a smart contract) in a past block, even if those values have later been updated. This functionality allows for an efficient verification of transactions of block $n$: when applied on the state of block $n-1$, these transactions, if valid, will update the state to that of block $n$. To implement this functionality in a space-efficient fashion, Ethereum uses an interesting variation of the copy-on-write technique, as illustrated in Figure~\ref{fig:memory}. While the transaction and receipt tries are strictly separate for each block, the state and storage tries can be partially shared across the blocks. In the illustration, the key for the state record of \emph{Account 1} in \emph{Block 1} hashes to value \emph{b10a5}. The record is placed in \emph{Leaf 1}, which is the leftmost child of the branch vertex. Similarly, the key for the state record of \emph{Account 2} in \emph{Block 1} hashes to value \emph{b1d14} and it is placed in \emph{Leaf 2A}. The state of \emph{Account 1} has not been modified by the transactions of \emph{Block 2}. Therefore, the branch vertex in \emph{Block 2} points to the same record of \emph{Leaf 1} as the branch vertex in \emph{Block 1}. On the other hand, the state of \emph{Account 2} has been modified by the transactions of \emph{Block 2}. 
Since the path to the state of \emph{Account 2} is the hash of the account address i.e., \emph{b1d14}, Ethereum creates a new record called \emph{Leaf 2B} in the same path under the branch vertex in \emph{Block 2} to update the balance. Due to the creation of a new leaf record, Ethereum also creates new branch and state root vertices in \emph{Block 2}. Note that by starting from the old root vertex in \emph{Block 1}, we can still find the old state of \emph{Account 2}.

Storage tries are updated in a similar way. As a result, the state and storage tries are only partially updated due to the transactions in every new block. In reality, the changes in the state and storage tries of each Ethereum block are fairly small~\cite{pruning}, which makes the copy-on-write technique described above particularly effective.

\subsubsection{In-memory storage}
\label{sec:in-memory}

Bitcoin creates Merkle trees in memory on demand, when a need arises (see Section~\ref{sec:tries}).
As mentioned in Section~\ref{sec:tries}, the trees are deterministically reproducible in memory from transactions stored on disk. 
%
%
Furthermore, in Bitcoin, the entire LevelDB database of block indexes  is loaded into memory on startup to achieve a better performance of finding the blocks~\cite{bitcoinStorage}. UTXO cache is another critical data structure maintained in Bitcoin memory~\cite{bitcoin-cache}. As the UTXO set stored in LevelDB is accessed frequently during block validation, the access time becomes a major bottleneck. Therefore, during the block validation, the required unspent coins are pulled from LevelDB into an in-memory cache. Since the size of the entire UTXO set exceeds 8GB, cache management is required.

In Ethereum, different node implementations such as Geth maintain an in-memory cache of configurable size, designated for vertices of various tries. In the context of Hyperledger Fabric, when using a state database, read delays during endorsement and validation phases have historically been a performance bottleneck. With Fabric v2.0~\cite{fabricCache}, a new peer cache replaces many of the expensive lookups with fast local cache reads, and the cache size can be configurable. Moreover, what is stored in the memory storage of Hyperledger Fabric depends on the implementation of the application. 
Finally, IOTA uses Tangle Accelerator~\cite{tangle-accelerator}, which is an intermediate caching proxy server between the client and IOTA node to speed up the attachment process of a bundle to tangle.

\subsubsection{Data Retention}
\label{sec:retention}

Data retention is performed differently in each of the representative blockchain systems. The ledger in Bitcoin grows more slowly compared to Ethereum and IOTA because Bitcoin stores less data per-block compared to Ethereum and because new blocks are appended less frequently compared to ledger changes in Ethereum or IOTA. As a result, Bitcoin is less in need for ledger pruning mechanisms. For the time being, Bitcoin does not have any data retention policy and keeps all of its transactions forever. As a consequence, the storage on Bitcoin full nodes can potentially grow large in size over time, which may become an even bigger issue if Bitcoin manages to overcome its scalability barrier and raise the average rate of transactions.

Similar to Bitcoin, Ethereum keeps its transactions forever in the ledger. However, Ethereum has a state trie pruning mechanism~\cite{pruning} for obsolete states. This mechanism tracks when trie vertices are no longer referenced by the state trie (for example, the \emph{Leaf2A} vertex in Figure~\ref{fig:memory} has dropped out of the state trie in \emph{Block 2}), and at that point places the dropped vertices on a death row in the database. From this point, after 5000 new blocks have been appended, the vertex will be permanently deleted from the database. Essentially, Ethereum stores the trie vertices that are part of the current state or part of the recent history~\cite{pruning} (up to 5000 recent blocks). 

Similar to Bitcoin and Ethereum, transactions on the ledger will never be deleted in Hyperledger Fabric. However, data from the worldstate database can be deleted if an application developer defines the procedure of pruning obsolete data in the implementation.  

IOTA provides snapshots regularly in order to reduce the size of storage needed by nodes. The snapshot process removes all the transaction history and the addresses with zero balances and creates the list of addresses with a non-zero balance to reduce the Tangle size~\cite{snapshot}. This in turn speeds up the time required to do synchronization among nodes. Another advantage is that the snapshot can indirectly improve the throughput of transactions in IOTA by increasing the speed of computing the balance for the accounts. Considering the nature of permanodes (refer to Section~\ref{sec:iota}), the snapshot process does not have any impact on their storage. 
\begin{table*}
	\setlength\tabcolsep{3pt} \setcellgapes{3pt} \makegapedcells
	\centering
	\caption{Communication layer of different blockchain systems}
	\label{tab:communication}
	\scalebox{1}{
		\begin{tabular}{|Y{2.8cm}|Y{3.3cm}|Y{3.5cm}|Y{3cm}|Y{2.9cm}| @{}}
				\hline
				\textbf{Features} & \textbf{Bitcoin} & \textbf{Ethereum} & \textbf{Hyperledger Fabric} & \textbf{IOTA} \\
				\hline
				Granularity of dissemination &  Whole network & Whole network & Per channel & Whole network\\ 
				\hline
				Entities forming the network &  Full nodes &  Full nodes & Orderers and peers & Nodes \\ 
				\hline
				Communication protocol & Inventory \newline push-gossiped \& blocks pulled by full nodes & Blocks or block hashes push-gossiped \& block headers and bodies pulled & leader pulls blocks from orderers \& push-gossips to peers & Push-flooding\\
				\hline
				Mean time to \newline receive a block & About 12.6 seconds & About 109 milliseconds & Application-dependent & No studies conducted \\ 
				\hline
				Ordering \newline guarantees &  No guarantees & No guarantees & Totally ordered \newline broadcast & No guarantees \\ 
				\hline
				Privacy \& \newline security \newline guarantees & No guarantees & Proprietary deployment can enable encrypted \& authenticated messages & Authenticated channels &  Authenticated and encrypted messages, partial anonymity \\
				\hline
				Initial peer \newline discovery &  Through a set of DNS seeds or direct conn. to a known full node & Through a set of \newline bootnodes or direct conn. to a known full node & Anchor peering \& \newline channel membership & Manual conn. through a node list \\			    
				\hline
				Geo-proximity in the network & 135 ms (avg.) peer-to-peer latency & 171 ms (avg.) peer-to-peer latency & Depends on consortium network topology & No studies conducted \\
				\hline
		\end{tabular}}
	\end{table*}

\subsection{Communication Layer}
\label{sec:communication}

Comparison between the communication layer features of the surveyed blockchain systems is shown in Table~\ref{tab:communication}. Only full nodes are part of the blockchain network in Bitcoin; they receive all blocks and transactions by running the dissemination protocol.  Lightweight nodes connect to the full node of their choice. 
Unlike full nodes, lightweight nodes only receive a subset of transactions, filtered for them by the
full nodes to which they are connected~\cite{bloom}. Furthermore, lightweight nodes rely on full nodes for propagating transactions to the network. In terms of receiving blocks, lightweight nodes download only the block headers from the full nodes.
Ethereum is absolutely identical to Bitcoin in terms of network formation and the division between full and lightweight nodes. Fabric and IOTA are also quite similar in this regard: peers constitute the peer-to-peer network in Fabric while clients connect to the peers which selectively relay transactions and blocks from and to the clients. In IOTA, clients connect to the network of IOTA nodes, the latter relaying bundles from and to the clients.

In order to manage the \emph{granularity of dissemination}, Hyperledger Fabric is using the concept of channels. Transactions submitted and propagated in a channel are isolated from the other channels. The channels are independent of each other in terms of the order, delivery, and processing of the transactions. On the other hand, transactions in Bitcoin, Ethereum and IOTA are propagated through the whole network.

Next, we consider the \emph{communication protocol} and \emph{propagation time}
for each of the four systems.
In Bitcoin, each full node sends inventory messages (\textit{inv})~\cite{bitcoinmessages} periodically to advertise its knowledge of transactions or blocks to the neighbors in the network. This advertisement contains the hash value of the transactions or blocks as their identifiers, which is much shorter than the actual transaction or block. Then, each neighbor that receives the advertisement checks if it mentions knowledge of transactions or blocks the neighbor has not seen before. In this case, the neighbor requests the missing transactions or blocks from the sender via the \textit{getData} message~\cite{bitcoinmessages}. 

Experiments in~\cite{propagation} indicate that it takes about 40 seconds for a new block to propagate to 95\% of the Bitcoin network (the mean time to receive a block for a full node is 12.6 seconds). This time includes the transmission time (for \textit{inv} message, \textit{getData} message, and delivery), as well as the verification time. The propagation time is likely to be even shorter for the transactions compared to blocks.

In Ethereum, a miner that creates a block or another full node that receives a new block from the network use the same propagation protocol. Namely, the full node considers a set $S$ of its neighbors and picks a random subset $S' \subset S$ of size $|S'| = \sqrt{|S|}$. Then, the full node forwards the entire block to full nodes in $S'$ while only sending a hash of the block to full nodes in $S \backslash S'$.
The neighbors receiving the hash of a new block request the block header via a GetHeader message~\cite{ethna} from either the sender or any of their neighbors that possess the block. Finally, after getting the new block header, the Ethereum full node requests the block body in a way similar to requesting the header.
 
The study of \cite{ethereum-propagation} shows that 95\% of the blocks are propagated through the Ethereum network within 211 milliseconds (the average time of a block propagation delay is stated to be 109 milliseconds). An experiment in  \cite{ethna} estimates that it takes only 3 or 4 hops to broadcast a new block to the entire Ethereum network. This small world of an Ethereum network compared to the Bitcoin network can be the reason for the lower block propagation time in Ethereum.

Peers in Hyperledger Fabric use both push and pull methods in the communication protocol~\cite{fabricgossip}. 
In the case of block dissemination, a leader is selected from the peers. The leader becomes responsible for pulling blocks from the orderers and then initiating a push-gossip protocol to propagate the blocks to the peers: Each peer broadcasts the blocks to a random set of neighbors in the channel. In addition to push-gossiping, each peer is also responsible for selecting a number of random peers regularly and attempting to pull the missing blocks from them. The design choice to use a leader for initiating push-gossip is motivated by the need to reduce the load of sending blocks from the orderers to the network. Orderers in Hyperledger Fabric also enforce basic access control for channels, restricting who can read and write data to the channels, and who can configure them~\cite{orderingservice}. 
In Fabric, the propagation time is dependent on application characteristics, especially on the transactions type and network size.

In IOTA, the transactions are flooded throughout the network similarly to Ethereum. On the other hand, since no studies have been conducted for the propagation time in IOTA, it can be considered an interesting subject for future research.

Regarding the \emph{ordering guarantees}, Hyperledger Fabric is the only system among the blockchain systems covered in this survey which supports total-order broadcasting~\cite{totalorder} within each communication channel. This guarantee ensures that different endorsers and orderers receive transactions from each given client ordered in the same way relatively to the transactions from other clients. The relative order of transactions is important because of the effects that the transactions may have on each other.

To support \emph{privacy and security guarantees}, transferred transactions between the Ethereum nodes are encrypted and authenticated~\cite{decentralization}, which makes the Ethereum communication layer more secure compared to Bitcoin. Furthermore, Whisper~\cite{whisper}, a configurable messaging protocol, which can be enabled in proprietary Ethereum deployments, provides the benefit of hiding the location of sender and receiver in the network, if such privacy requirements arise. Whisper can achieve this benefit by using onion routing and other techniques common in the Tor~\cite{tor} network.
Hyperledger Fabric utilizes the authenticated channels to provide the privacy and security of the network.
 IOTA has also introduced the functionality of issuing and accessing encrypted data streams by implementing the Masked Authenticated Messaging (MAM) protocol~\cite{mam} as a second layer communication protocol to improve the privacy and security of the communication layer.



\begin{table*}
	\setlength\tabcolsep{3pt} \setcellgapes{3pt} \makegapedcells
	\centering
	\caption{Data manipulation layer of different blockchain systems}
	\label{tab:manipulation}
	\scalebox{1}{
		\begin{tabular}{|Y{2.5cm}|Y{2.8cm}|Y{3cm}|Y{3.7cm}|Y{3.4cm}| @{}}
				\hline
				\textbf{Features} & \textbf{Bitcoin} & \textbf{Ethereum} & \textbf{Hyperledger Fabric} & \textbf{IOTA} \\
				\hline
				Consensus  protocol & Chain convergence using longest-chain rule  & Chain convergence using GHOST  protocol & Configurable consensus module (Raft is the default) & DAG convergence using tip selection algorithm \\		  	  
				\hline
				Mining difficulty & About 10 minutes to mine a block & About 12-15 seconds to mine a block & No mining & No mining \\ 
				\hline
				Throughput (tps) & 3-7 & About 15 & Configuration-dependent & About 50 \\ 
				\hline 
				Mining power \newline utilization & Above 99\% & Below 97\% & No mining & No mining \\ 
				\hline  
				Transaction \newline confirmation & Probabilistic based on the number of blocks & Probabilistic based on the number of blocks & Deterministic, \newline after committing on the peers & Probabilistic based on confirmation confidence  \\
				\hline
				Mitigating DoS \newline attacks & PoW & PoW + Gas price & Permissioned \newline authentication & PoW\\
				\hline
				Rich search \newline functionality & No & No & SQL-like query & No \\ 
				\hline
		\end{tabular}}
	\end{table*}

\emph{Geographical proximity} between participants in the blockchain network is one of the parameters that have effect on the block propagation time throughout the network. Measurements in~\cite{decentralization} indicate that Bitcoin exhibits a higher degree of clustering compared to Ethereum, i.e., it has many more groups of full nodes that are geographically close to each other. The measurements also show that an estimated peer-to-peer latency between Ethereum full nodes is 26.7\% higher on average compared to Bitcoin. The main reason for this higher clustering degree is that many Bitcoin full nodes are running in data centers. This research claims that 56\% of Bitcoin full nodes belong to autonomous systems that provide dedicated hosting services whilst the percentage for Ethereum full nodes is 28\%. On the other hand, in permissioned blockchains such as Hyperledger Fabric, geographical proximity between peers depends on the consortium network topology. To the best of our knowledge, geographical proximity between IOTA nodes has not been studied yet.

\subsection{Data Manipulation Layer}
\label{sec:datamanipulation}


Data manipulation in blockchain primarily consists of two operations: updating the state of the blockchain and querying it. Since the transactions stored in the blockchain are immutable, the only update operation allowed is adding a new transaction to the ledger, which additionally results in performing bookkeeping needed to keep track of the updated state. While the goal is simple, the update process is quite complex because it requires running a consensus protocol across the participants in the blockchain network. The blockchain type and even design goals vary across the systems, which results in significant variations in the flavor of consensus protocols used~\cite{Marko,Marko2}.  This is the reason why blockchain systems exhibit significant variations in their performance and in security assumptions and threats they can withstand. Besides, the performance is affected by deployment characteristics, such as the blockchain network, transaction rate, duration of transaction execution, and the distribution of mining power in systems that use mining. Accordingly, most of this section is dedicated to the update protocol in the four systems considered. For each system, we also discuss the security and performance properties. Table~\ref{tab:manipulation} shows the differences in the features of the representative blockchain systems in the data manipulation layer.

Querying the blockchain is conceptually much simpler than updating it because it does not require coordination across the participants in the blockchain network and because the scalability and security constraints are not nearly as severe. However, the differences in blockchain types and stored data (see Section~\ref{sec:datastorage}) lead to different query models.

Finally, security plays a very important role in the design of the data manipulation layer. Yet, different blockchain systems exhibit differences in the security assumptions and tolerated threats so that there exist no easily identifiable baseline to compare blockchain systems against. For example, the issues of branching and concentration of computing power in the hands of a single participant (or a group of colluding participants) is a major consideration for Bitcoin and Ethereum. However, it is a non-issue for Hyperledger Fabric which does not use PoW and where the branching is impossible due to the consensus protocol being deterministic.
The security model behind IOTA has only been partially explored and defined. Accordingly, we discuss security aspects of each system separately in the context of presenting the consensus protocol. However, all blockchain systems must be resilient to denial-of-service (DoS) attacks, which is why we include a focused discussion comparing the defense mechanisms to DoS attacks in all four systems.

%


\subsubsection{Updating blockchain in Bitcoin}
\label{sec:manipulation-bitcoin}

The mechanism of updating the blockchain in Bitcoin has been presented in other survey works (see Section~\ref{sec:relatedWorks}) and in a textbook~\cite{bitcoinText}. In this section, we provide an extended description written in a pedagogical way. While containing limited novelty by itself, this description serves as a basis for comparison with the more novel descriptions of updating the blockchain in the other three systems, presented in the later sections.

The general flow of updating the blockchain can be found in Section~\ref{sec:bitcoin}. What is especially interesting about the blockchain update solution in open blockchain such as Bitcoin is that it is fully self-moderated: the nodes are fully autonomous. They do not have to follow the complex protocol but do so because they are given a combination of incentives and protective mechanisms. For example, the consensus protocol is very sensitive to the rate of proposals: too many and the protocol stops to scale; too few and the system does not make any progress. 

The PoW mechanism prevents the miners from issuing proposals at too high of a rate. 
The rate is regulated by the complexity of the cryptopuzzle, called \emph{mining complexity}. Bitcoin adjusts the mining complexity every 2016 blocks so as to keep the average interval between consecutive block proposals limited to 10 minutes. To be precise, every miner in Bitcoin adjusts its own mining complexity independently based on its local copy of the chain. However, since the adjustment is based on the height of a block and not on the data in the chain,
every miner performs the same adjustment after the same block number, which ensures fairness of mining.



In order to keep the rate up, Bitcoin incentivizes miners to spend their resources on solving cryptopuzzles. Specifically, it introduces two incentive mechanisms: (a) a block creation reward that is given to a miner whose block has been successfully included into the chain, and (b) a transaction inclusion fee that a transaction creator may voluntarily pay to the miner that includes the transaction into the proposed block. The block creation reward is regulated by Bitcoin similarly to the mining complexity. It is halved every 210000 blocks. Taking into account the average 10 minute interval between consecutive blocks, the reward is expected to become zero before 2140. Currently, the block creation reward plays a more important role than a transaction inclusion fee but as the reward gets smaller, the transaction inclusion fee will start playing an increasingly more important role. Every block includes a single transaction assigning block creation reward to the proposer; this is the only mechanism by which Bitcoin injects new funds into circulation. In order to increase the chance of receiving block creation reward, a group of miners can organize themselves into a  \emph{mining pool}. Then, all members of the pool work together to mine each block, and the revenue of mining a block is shared among them. 

\begin{figure}
		\includegraphics[width=\linewidth]{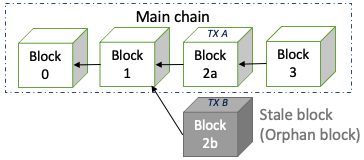}
		\caption{Bitcoin chooses the main chain based on the longest chain rule. Any valid block outside the main chain will be ignored and called a stale block.}
\label{fig:stale}
\end{figure}

In Bitcoin, there is no entity in the system that knows the precise state of the updated global chain. Every full node is trying to approximate the updated knowledge of the global chain by maintaining a local copy of it (ledger and UTXO information as explained in Section~\ref{sec:datastorage}); the full node may update the local copy upon receiving a new block proposal. 
As different Bitcoin miners may create and propagate different valid block proposals in parallel, the local ledger copies on different full nodes may temporarily diverge.
In the example of Figure~\ref{fig:stale}, Block 2a and Block 2b are both valid and created in parallel as successors to Block 1. This situation when different blocks are referring to the same predecessor block is called a \emph{fork} in the blockchain. 
In addition to the accidental fork creation which happens because of concurrent block proposals, there are also intentional forks. Whenever the blockchain protocol is updated (for example when the block size is changed) on a subset of nodes, it can result into two types of fork: hard and soft. In the hard fork situation, the nodes with updated protocol cannot interoperate with the nodes running the old version of the protocol at all. A hard fork can lead to creation of a new cryptocurrency. For example, Bitcoin Cash~\cite{bitcoincash} has been created as a hard fork of Bitcoin, and is now completely independent of Bitcoin. On the other hand, soft forks support backward compatibility. It means that transactions from nodes running the old protocol are accepted by the nodes running the new one, but transactions created by the new protocol addresses are not accepted by the old nodes. Segregated Witness (SegWit)~\cite{segwit} is an example of a soft fork. SegWit is created as a protocol upgrade of Bitcoin to improve block capacity and prevent the transaction malleability (changing a transaction after signing it) problem by introducing a new transaction structure. 

In the case of accidental fork creation, the miners have to decide which of the branches will be included in the main chain and which branch will be discarded. In order to reach a consensus on the included branch, Bitcoin has the longest chain rule for resolving conflicts between competing branches~\cite{nakamoto}. This rule stipulates that the longest branch wins and becomes part of the main chain for building the subsequent block upon. The rationale for this rule is that the longest chain corresponds to the biggest amount of work on the cryptographic puzzles; by including this branch and discarding shorter branches, we are wasting the least amount of work. Eventually, the local copies converge to the same global state.

In the example of Figure~\ref{fig:stale}, Block 3 points to Block 2a as its predecessor while ignoring Block 2b. This may happen because the miner creating Block 3 received 2a before 2b.
In this case, Block 2b is discarded according to the longest chain rule despite being valid and verified. It is called a \emph{stale} (sometimes called \emph{orphan}) block~\cite{orphan}. 

It should be noticed that because the convergence of the local copies is only eventual, this complicates the reasoning about the blockchain state. 
A full node may attempt to infer information about the global state based on the local copy, without any consultation with other full nodes. The most common case of inference is when a full node tries to establish whether a given transaction has been included in the blockchain. The inference is correct with a certain probability, which never reaches 100\%: A transaction $A$ may be included in a block and added to the local copy, but at a later time, a different branch may become the longest and win the race, in which case $A$ will not be included in the chain. However, the more subsequent blocks are added to the local copy after the block $B_A$ containing $A$, the longer the branch in the local copy becomes, which reduces the probability that a competing branch not including $B_A$ will become longer yet. In Figure~\ref{fig:stale}, transaction $A$ (\emph{TX A}) and transaction $B$ (\emph{TX B}) are included in Block 2a and Block 2b respectively. The inclusion probability of transaction $A$ in the main chain is increased by creating Block 3 upon Block 2a, while, the chance of inclusion in the chain for transaction $B$ is reduced since it is not included in the longer chain. A more precise calculation of probabilities for the finality of transaction inclusion can be found, e.g., in~\cite{confirmationBitcoin}. Bitcoin full nodes use this reasoning to provide a confirmation about the inclusion of transactions: after 6 subsequent blocks are included into the local copy, the full node considers the probability high enough to send a confirmation to the user. For example, a transaction may transfer a fee in bitcoins to a service provider. 


However, a malicious miner that owns more than 50\% of the network mining power, may overcome the network by mining, producing valid blocks, and building longest chains faster than the rest of the network. In particular, it gains the ability to tamper with the data and subvert the blockchain.
This situation is called a 51\% attack. The miner performing the attack can exploit the situation in different ways, such as ignoring some specific transactions or depriving the competing miners from receiving block creation rewards. It can also define a new policy and impose it onto the network. As an example, the exact amount of block creation reward is configurable and subject to an agreed policy. An attacker will be able to increase the block creation reward in hope that the attacker will be mining most blocks and ripping the benefits of a higher reward. Furthermore, the attacker that owns more than 50\% of the network mining power can perform a \emph{double-spending} attack, that is to transfer the same virtual coin in two different transactions. For example, an attacker creates one transaction to send some bitcoins to a recipient in exchange for a service or product. This transaction is included in Block 2b, which is appended to the blockchain. Let us denote the predecessor of Block 2b as Block 1 (see Figure~\ref{fig:stale}). After receiving the service or product from the recipient of the transaction, the attacker creates another transaction that sends the same bitcoins to a different address of his own. The powerful attacker quickly creates and mines Block 2a with the new transaction, and then creates and mines Block 3 as a successor of Block 2a, in order to ensure that Block 2a is included in the longest branch, and that Block 2b becomes stale. As a result, the attacker gets the service or product without paying any bitcoins for it.

The probability of a 51\% attack being successful depends on a number of factors~\cite{ghost}, above all those contributing to the presence of forks. The situation illustrated in Figure~\ref{fig:stale} is undesirable from the security point of view. This is because Block 2b does not contribute to securing the main chain so that the attacker requires less mining power to subvert the chain. To capture this point, \cite{bitcoin-ng} has introduced the metric of mining power utilization, which is defined as the ratio between the mining power that secures the main chain and the total mining power.  Mining power wasted on work that does not appear on the blockchain accounts for the difference. The higher the mining power utilization, the more mining power an attacker must have in order to subvert the chain.

The argument of mining power utilization is the main reason against reducing the complexity of PoW in Bitcoin. Even though such a reduction may improve the throughput of transactions and reduce transaction confirmation time (that is the time required for a transaction to reach a certain probability threshold for being included in the chain), it will lead to significant forking and reduced security~\cite{ghost,prism}.

The subject of block size is a hotly debated topic in Bitcoin where no common view has been reached. Original Bitcoin design by Satoshi Nakamoto has introduced a limit on the maximum block size; this limit still applies. A bigger block size would allow more transactions to be included in the block, thereby improving the throughput of transactions. On the other hand, a bigger block would potentially increase the block propagation time (see Section~\ref{sec:communication}), which may give an advantage to the miner of the current block who can start mining the next block earlier before it is disseminated to other miners. This may potentially increase the centralization in the Bitcoin network. However, the exact effect of the latter is very difficult to capture and analyze without deployment, which contributes to the controversy; some believe that a limit on the block size is artificial and can be relaxed or even eliminated.

An attack strategy called \emph{selfish mining}~\cite{selfish} was proposed to show that even with less than half of the network's mining power, misusing of the Bitcoin's longest chain rule is possible. The principal claim of this work is that a selfish mining pool with 1/3 mining power of the network can still defeat the honest mining protocol, and the revenue of a selfish pool rises superlinearly as the pool size grows. However, some researches such as~\cite{fallacy} and~\cite{whynotselfish} refute selfish mining strategy's assumptions and benefits, and claim that there is no advantage for a selfish miner to follow the attack strategy instead of the honest one. Selfish mining has not been observed in the Bitcoin network in practice which reinforces these claims. Even if a selfish mining attack succeeds in the short term, it may lower the value of Bitcoin, which further reduces the benefits of the attack compared to honest mining.

\subsubsection{Updating blockchain in Ethereum}
\label{sec:manipulation-ethereum}

Conceptually, the protocol of updating the blockchain in Ethereum is similar to that of Bitcoin: Both systems are permissionless and use mining. Both systems use the same entities that play the same roles, e.g., the miners propose new blocks of transactions. In both systems the full nodes maintain local copies of the ledger which eventually converge but may temporarily diverge.

At the same time, the two systems exhibit significant differences related to the design goals and choices, and deployment characteristics, as we elaborate below. These differences affect both performance metrics (such as transaction throughput, transaction confirmation time and chain convergence speed, and energy consumption) as well as system security and the risk of centralization.

With respect to design goals, Ethereum supports general transactions written in a Turing-complete programming language. This means that the execution of transactions themselves is more resource- and time-consuming in Ethereum compared to Bitcoin. This leads to reduced throughput and increased energy consumption. Since executing transactions is unlikely to be as computationally intensive as solving cryptopuzzles, the effect is likely to be less pronounced compared to that of PoW. However, it has never been analyzed. 
Another interesting aspect of Turing-complete transactions is that it becomes absolutely essential to limit a block. 
However, unlike in Bitcoin, it does not make sense to express the limit in bytes because the block size is no longer the main limiting factor. Instead, the limit is specified in the amount of computation required, which is expressed in the maximum amount of Ethereum gas consumed. Of course, the limit on the amount of computation also indirectly restricts the length of transactions and their number in a block and thereby, the block size. In September 2019, the average block gas limit was around 10,000,000 units~\cite{Conner-2019}, which translated to the average block size of between 20 to 30 kilobytes~\cite{Ethereum-Block-Size-Chart}, based on the transaction rate and transaction computational requirements in Ethereum at that point in time.

Another difference in design goals is that Ethereum supports more complex state storage (see Section~\ref{sec:state-tracking}), which results in bigger state storage sizes compared to Bitcoin. While this has an impact on most performance metrics, the exact effect has never been scientifically studied. 


Regarding design choices, Ethereum uses a somewhat different conflict resolution rule and a block reward model, as well as less complex cryptopuzzles. Besides, a higher number of subsequent blocks is required for transaction confirmation in Ethereum compared to Bitcoin. We now consider these in detail.






For conflict resolution, Ethereum is using a modified version of Greedy Heaviest Observed Subtree (GHOST) protocol~\cite{ghost} instead of the longest-chain rule. This fact is mentioned in~\cite{ethereumWhitePaper} without comparing the original version of the GHOST protocol with the version used in Ethereum. We believe that our analysis below provides the first complete comparison in the literature. 

\begin{figure}
	\includegraphics[width=\linewidth]{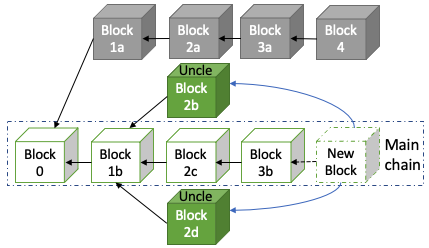}
	\centering
	\caption{Ethereum chooses the main chain based on the GHOST protocol. Uncle blocks are considered when choosing the main chain (the heaviest chain rule) and their miners are rewarded in Ethereum.}
	\label{fig:uncle}
\end{figure} 

\begin{table*}
	\setlength\tabcolsep{3pt} \setcellgapes{3pt} \makegapedcells
	\centering
	\caption{How differences from Bitcoin affect various performance and security metrics in Ethereum}
	\label{tab:factors-metrics}
	\scalebox{0.95}{
		\begin{tabular}{| l | l | l | l | l |} 
			\hline
			\textbf{Factor} & \textbf{Throughput} & \textbf{Confirmation time} & \textbf{Energy consumption} & \textbf{Decentralization and Mining power utilization} \\
			\hline
			General transactions & worsens & worsens & worsens & worsens  \\
			\hline
			Complex storage & worsens & worsens & worsens & unknown  \\
			\hline
			Considering uncles & unknown & unknown & unknown & improves  \\
			\hline
			Less complex cryptopuzzles & improves & improves & improves & worsens \\
			\hline
			Deployment characteristics & unknown & unknown & unknown & unknown  \\
			\hline
	\end{tabular}}
	\vspace{-3mm}
\end{table*}

The first point of departure in all versions of the GHOST protocol compared to Bitcoin is that the miner producing a block is incentivized to reference a limited number of stale blocks, in addition to the previous block. In Ethereum, the miner can specifically reference up to two stale blocks from the header of the newly generated block, as shown in Figure~\ref{fig:uncle}. The miner is incentivized to do so by increased block creation reward for each referenced stale block. A stale block that is not included in the main chain but referenced by a main chain's block is called \emph{uncle} or \emph{ommer} block in Ethereum. In the example of Figure~\ref{fig:uncle}, Blocks 2b and 2d are uncle blocks for the new block. The uncle block's miner is also compensated, so that the miner's effort of producing the block does not go unrewarded.

Secondly, both the original GHOST protocol and the version used in Ethereum consider the uncle blocks and merge the amount of work that has been done for each block in different branches in order to choose the main chain. For example, in Figure~\ref{fig:uncle}, the main chain is chosen based on the GHOST protocol, and the result is different from the longest chain rule. The algorithm starts from Block 0 and at each fork, chooses the block leading to the heaviest subtree. In the given example, the subtree of Block 1a contains 4 blocks, while the subtree of Block 1b (prior to the inclusion of the new block) consists of 5 blocks. Assuming that each block has the same cryptopuzzle complexity, Block 1b will be chosen. According to the same rule, Blocks 2c and 3b are included in the main chain. Then, a miner that wants to propose a new block, will choose Block 3b as the predecessor. 

The original GHOST protocol additionally considers transactions in uncle blocks when computing the blockchain state. This element is not present in Ethereum.

Since the stale blocks are adding to the weight of the main chain, forks are less affecting mining power utilization in Ethereum compared to Bitcoin~\cite{difficultyAdjustment}, which makes the main chain more secure and reduces the risk of centralization. As a consequence, Ethereum can afford shorter block creation time, which results in a higher number of forks but also better decentralization, transaction throughput, and confirmation time~\cite{ghost,prism}. Specifically, the average time to solve a cryptopuzzle is 12-15 seconds in Ethereum compared to 10 minutes in Bitcoin~\cite{ethereumthroughput}.

Regarding the confirmation time, an Ethereum full node confirms inclusion of a transaction after 12 subsequent blocks are included into the local copy of the blockchain~\cite{confirmationEthereum}, in contrast to 6 blocks in Bitcoin. The probability of a transaction inclusion being final has not been analyzed as thoroughly for Ethereum as it has been for Bitcoin. However, as the Ethereum's block creation time is shorter compared to Bitcoin's, Ethereum transactions are confirmed faster.

Table~\ref{tab:factors-metrics} presents a summary of the discussion in this section, with a particular emphasis on how each design difference in Ethereum compared to Bitcoin affects a variety of performance and security metrics. The comparison is based on the conceptual differences rather than empirical measurements. For example, the fact that transactions in Ethereum are Turing-complete programs may make them much more computationally intensive compared to simpler transactions in Bitcoin, resulting in worse throughput, higher energy consumption, and increased confirmation time. It does not mean, however, that actual Ethereum transactions are more computationally intensive in practice because the users may be unwilling to exploit the capabilities provided by Ethereum and pay the higher gas price of such transactions. The ``unknown'' entries signify research gaps: they indicate that it is difficult to assess the impact without further analysis and that the volume of studies analyzing the effect of the corresponding design element on a particular metric is limited.

Finally, the characteristics of the main Ethereum deployment, namely the Ethereum network, distribution of the mining power across the miners, and the rate and computational requirements of transactions have a non-trivial impact on the performance and security metrics. 
An empirical study of various metrics in Ethereum is available in~\cite{decentralization}. Such a study can only evaluate the actual performance of the system as a whole; it cannot evaluate the effect of each design element separately from the rest of the system. This work shows, e.g., that the mining power utilization in Ethereum is 97\% compared to 99\% in Bitcoin. Besides, the throughput in Ethereum is 15 transactions per second~\cite{ethereumScaling} compared to 7 transactions per second in Bitcoin~\cite{scaling}.

\subsubsection{Updating blockchain in Hyperledger Fabric}
\label{sec:manipulation-fabric}

The consensus protocol in Hyperledger Fabric is fundamentally different from those in Bitcoin and Ethereum. First, the entities are different as explained in Section~\ref{sec:hyperledger}. Second, Fabric provides the ability to partition a consortium network into independent channels so that each channel can execute a consensus protocol in parallel and independently of other channels. In a sense, there is a separate ledger maintained for each channel. Besides, only authorized peers can participate in the consensus for each channel, as determined by the consortium policies and related channel configuration.
Third, the consensus protocol in Fabric is deterministic unlike Bitcoin and Ethereum where there could temporarily be divergent branches of the ledger that eventually and probabilistically converge. In Fabric, there cannot be divergent branches and the transaction becomes final and 100\% confirmed relatively fast.

A different class of consensus protocols leads to a remarkable contrast in performance characteristics.
As discussed in Sections~\ref{sec:manipulation-bitcoin} and~\ref{sec:manipulation-ethereum}, Bitcoin and Ethereum can achieve throughput of 3--7 and 12--15 transactions per second respectively when updating the ledger. On the other hand, \cite{fastfabric} has shown that the throughput in Hyperledger Fabric can reach many thousands of transactions per second with specific optimizations in place, as we discuss in the later part of this section. While providing a high transaction rate, the deterministic consensus protocol in Fabric does not scale well w.r.t. the number of peers in any single channel. At the same time, the probabilistic consensus protocols of Bitcoin and Ethereum can handle the scale of many thousands of full nodes, which is beyond the reach for Fabric~\cite{Marko}. 



Ledger update in Fabric is performed in three phases called proposal, ordering, and validation~\cite{orderingservice}. In the proposal phase, clients send every transaction proposal to the endorsers (see Section~\ref{sec:hyperledger}), which are responsible for simulating the transaction without updating the database. The endorsement policy specifies which endorsers a transaction is sent to for approval, and how many endorsements are needed for a transaction. For example, a policy can specify that \emph{m} signatures out of \emph{n} endorsers are enough for a given transaction to proceed to the ordering phase. 

During the ordering phase, the orderer service runs a consensus protocol and decides on the order in which concurrent endorsed transactions will be executed and added to the ledger. Hyperledger Fabric uses pluggable modules for the ordering service which allows for significant configurability. Currently, the Raft~\cite{orderingservice} implementation based on the Raft protocol~\cite{raftprotocol} is the recommended option due to being easier to set up and manage compared to the alternatives~\cite{orderingservice}. However, PBFT~\cite{pbft}, Apache Kafka~\cite{kafka}, BFT-SMaRt~\cite{bftsmart}, SBFT~\cite{sbft}, and other implementations can also be used as the orderer service for Hyperledger Fabric. The choice of the implementation and specific consensus protocol may significantly affect the throughput of adding transactions to the ledger.

Additionally, the resilience of the entire ledger update procedure is mainly inherited through the fault-tolerant guarantees offered by the chosen ordering protocol. The two most popular families of protocols for implementing ordering are Paxos and View-stamped replication~\cite{replication}. Both protocols tolerate crash failures when the number of faulty processes is below $n/2$, where $n$ is the total number of participants.  To tolerate byzantine failures, ordering protocols such as BFT-SMaRt~\cite{bftsmart} or PBFT~\cite{pbft} can be employed~\cite{Marko2}. These protocols work when the number of faulty participants is below $n/3$.

Then, in the validation phase, the orderer service sends the block containing ordered transactions to the peers, and the peers validate the correctness of transactions. Transactions successfully validated and committed at this stage are never revoked, which means that their inclusion into the chain is final and deterministic.

It is interesting to observe that in Bitcoin and Ethereum, the block proposer (miner) broadcasts the blocks only after verifying that all transactions in the block are valid.  In contrast, the block proposer (orderer) in Fabric does not validate transactions while the blocks are created in phase two. Transaction validation only happens at the proposal (first) phase by the endorsers and at the validation (third) phase of the consensus protocol.
While this makes the protocol susceptible to denial-service of attacks by polluting the blocks with invalid transactions, the permissioned access allows Fabric to make an assumption that there is no significant incentive for the entities participating in the consensus protocol to launch a deliberate security attack. The design decision to offload the transaction validation task from the block proposer confers performance benefits in terms of higher throughput at the cost of ledger potentially storing invalid transactions.

The study of \cite{thakkar} shows a number of interesting performance characteristics of the Hyperledger Fabric as follows. 
First, there is a saturation point for throughput at around 140 transactions per second (tps). When the transaction arrival rate reaches the saturation point, the commit latency of blocks is increased from 100ms to tens of seconds. The raise in the latency is due to the increased number of ordered transactions waiting in the queue for the peers to validate their correctness in the validation phase. In other words, the validation phase becomes the bottleneck.  Furthermore, when the transaction arrival rate is lower than the saturation threshold, the increase in block size leads to the increase in the block creation time in the orderer phase which in turn impacts the transaction latency. Additionally, when the arrival rate is 50 tps, the increase in the block size from 10 to 100 causes transaction latency increase by a factor of five from 242ms to 1250ms. On the other hand, it has also been demonstrated that when the transaction arrival rate is higher than the saturation threshold, the increase in block size leads to decreased transaction latency and enhanced throughput. This is due to the fact that the time taken to validate and commit one large block is shorter than the time taken to validate and commit many small blocks.

\begin{figure*}
	\includegraphics[width=175mm,scale=1]{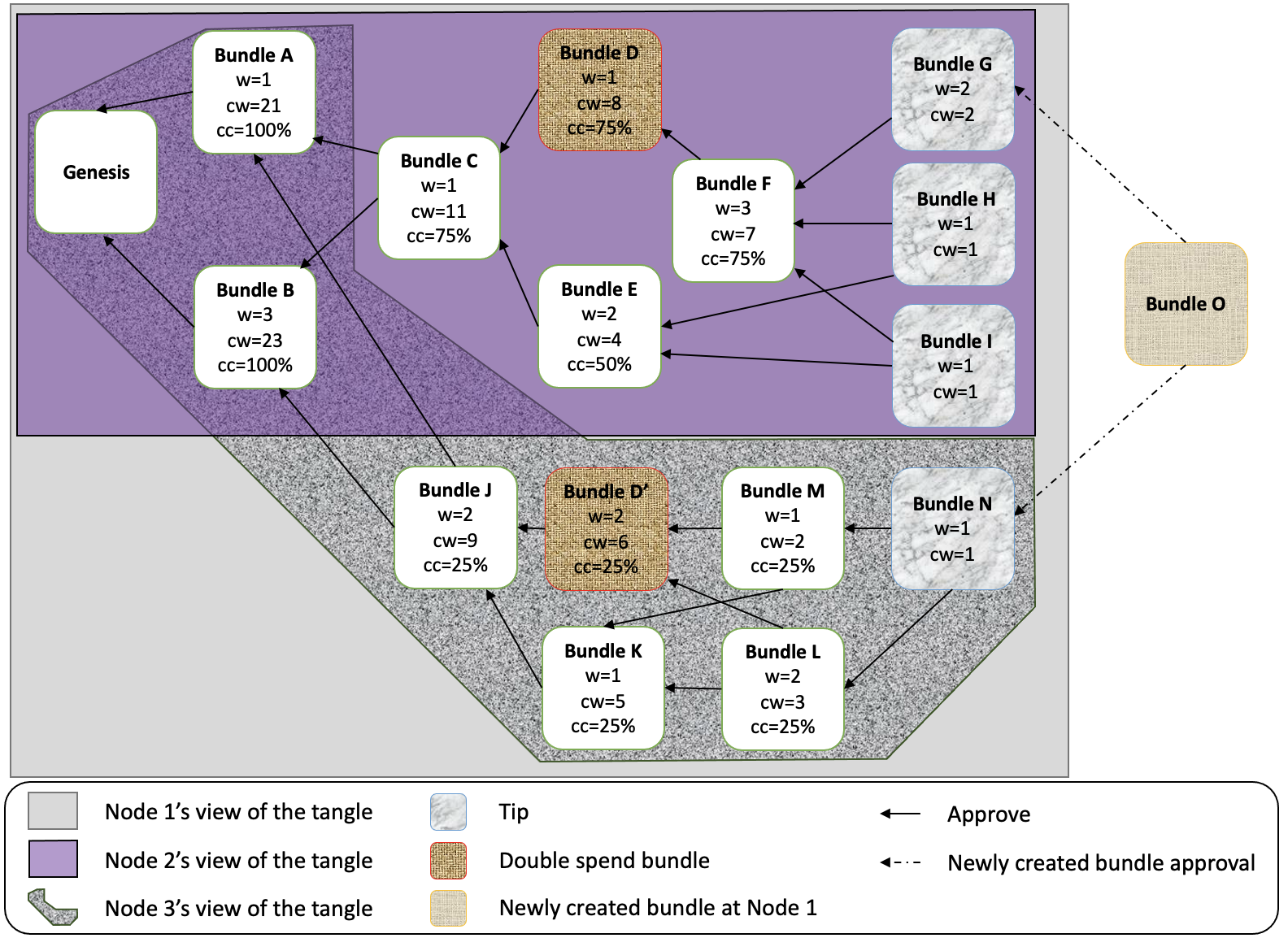}
	\caption{Every node has its own copy of the Tangle; copies might be diverging}
	\centering
	\label{fig:tangle-example}
	\vspace{-2mm}
\end{figure*} 

Additionally, it is demonstrated in~\cite{thakkar} that heterogeneity in the resources of peers and networks of organizations leads to performance degradation. Furthermore, this work shows that increasing the number of channels, using fewer endorsers, and performing bulk read/write operations contribute to better throughput and reduced latency in Fabric. 

Moreover, common system optimization techniques are used in~\cite{fastfabric} to achieve end-to-end transaction throughput of around 20,000 transactions per second. Examples of these techniques include separation of concerns (separating metadata from data while creating blocks in the ordering service), using parallelism and caching during the transaction validation, and replacing the database of worldstate with lightweight in-memory data structure  such as hashtable. One of the key assumptions in~\cite{fastfabric} towards achieving high throughput is that the incoming transaction workload is contention-free.  When there is a high contention and a large number of incoming transactions compete for a small set of hot keys in the worldstate, the throughput drops down significantly as discussed in~\cite{XOX}. 

The authors of~\cite{XOX} propose changes to the transaction flow in Hyperledger Fabric to handle both high and low contention transaction workloads. They show that the achieved transaction throughput (3000 transactions per second) is significantly better than in Fabric and~\cite{fastfabric} when the workload exhibits high contention. 
One of the key design ideas in~\cite{XOX} is to separate the dependent and independent sets of transactions in the block and process both sets concurrently during the validation phase. To realize this idea, a transaction dependency analyzer is introduced, whose goal is to isolate the transactions that are having overlap on the keys of their R-W sets with transactions that do not have any overlaps. As a consequence, the knowledge regarding the subset of transactions that can be processed concurrently is gained. The dependent transactions are marked invalid in the R-W set validation and will be re-executed on the latest version of the worldstate. The independent set of transactions that are marked valid can be successfully committed concurrently. 



\subsubsection{Updating blockchain in IOTA}
\label{sec:manipulation-iota}

\begin{figure*}
	\includegraphics[width=175mm,scale=1]{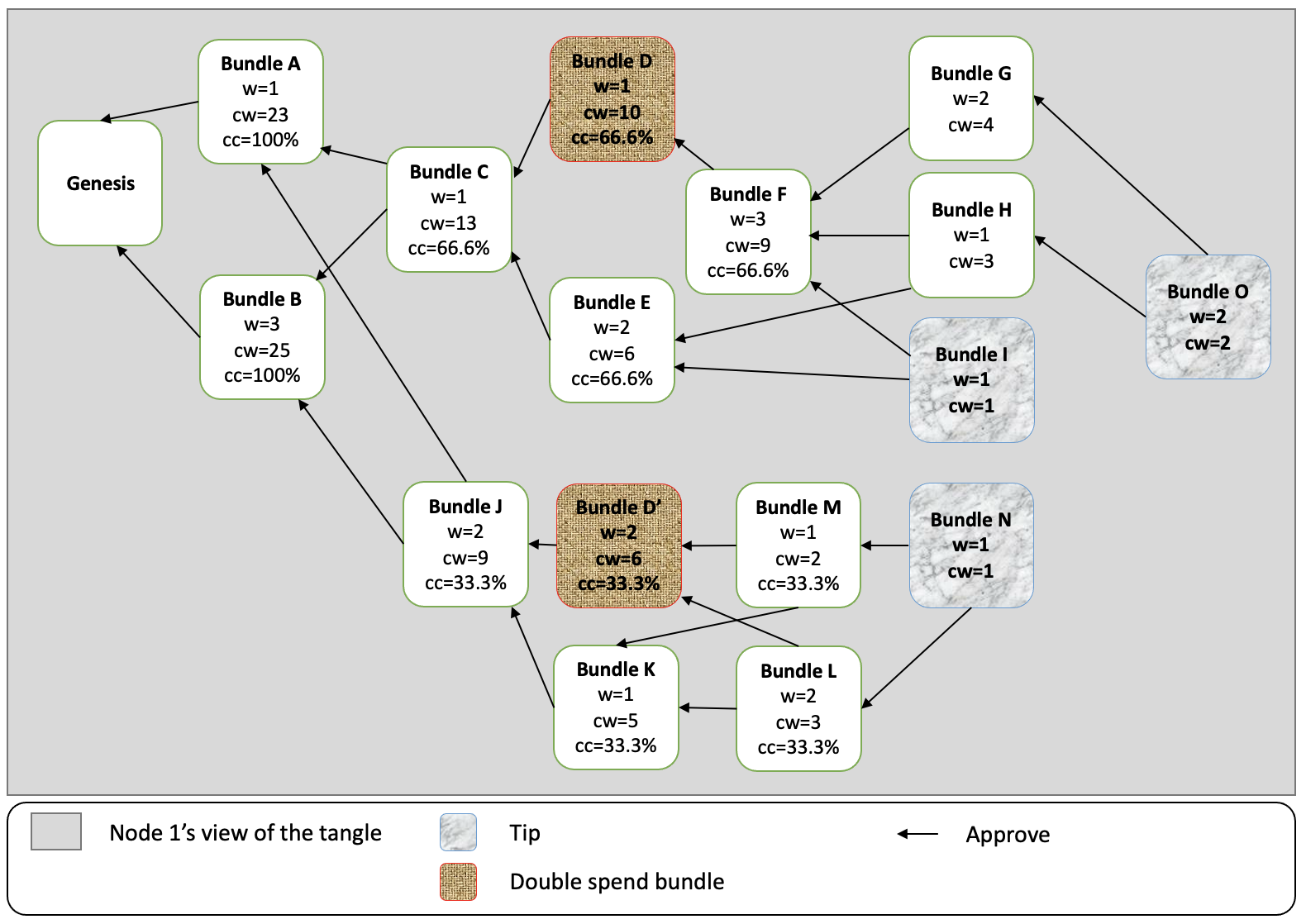}
	\caption{Choosing a pair of tips happens repeatedly until a valid pair is found}
	\centering
	\label{fig:tangle-example-reattached}
	\vspace{-5mm}
\end{figure*} 

While the consensus protocol in IOTA is probabilistic, it is different from the protocols in Bitcoin and Ethereum in many respects. Neither the concept of miner nor the chain of blocks exist in the IOTA network. 
%
%
As it is explained in Section~\ref{sec:datastorage}, IOTA utilizes a specially crafted DAG called Tangle instead of a chain of blocks. In a Tangle, vertices represent bundles (or transactions) while the edges signify approval of the bundles (see Figure~\ref{fig:tangle-example}). A Tangle starts with a genesis bundle and grows as new bundles are added. Each new bundle is supposed to validate and approve two bundles already included in the Tangle, though there are exceptional cases when only one bundle is validated and approved, instead of two. Similar to the ledger in Bitcoin and Ethereum, each node participating in the IOTA network maintains its own copy of the Tangle and synchronizes it with other nodes; these copies may be temporarily diverging. Figure~\ref{fig:tangle-example} illustrates a situation where participating nodes 2 and 3 have a partial knowledge of the Tangle while node 1 possesses combined knowledge.

In IOTA, every node $x$ that wants to issue a transfer of funds has to perform the following sequence of actions:
\begin{asparadesc} 
\item[Bundle creation:] 
$x$ creates a bundle including all input and output transactions that together constitute the transfer (see Section~\ref{sec:trans-struct}). For example, node 1 creates bundle $O$ in Figure~\ref{fig:tangle-example}.
\item[Weight assignment:]
$x$ decides on an integer weight $w$ for the bundle. A bigger weight results in a probabilistically faster confirmation for the bundle (as we explain below) at the expense of a more significant computational effort for $x$.
\item[Cryptopuzzle solution:]
$x$ needs to solve a cryptopuzzle for each of the bundle's transactions. The difficulty of the cryptopuzzle is proportional to $w$ selected by $x$. It is however, significantly lower compared to Bitcoin and Ethereum because in the latter, the cryptopuzzles regulate proposal creation rate while in IOTA they only prevent DoS attacks.
\item[Tip selection:]
$x$ selects two \emph{tips} to approve where the tip is a bundle already included in the Tangle but not yet approved by any other bundle. For examples, bundles $G$, $H$, $I$, and $N$  are tips in Figure~\ref{fig:tangle-example}. To this end, $x$ runs a tip selection algorithm~\cite{tangle} twice, once for each tip. We describe the tip selection algorithm in detail below. In addition to the tip itself, the algorithm selects a path from the genesis bundle to the tip. 

In exceptional cases, only one tip exists in the local copy of the DAG maintained by $x$ when $x$ creates a new bundle. For example, when node 3 created bundle $K$, only a single tip $J$ was available in the DAG. In such cases, only a single tip is selected and approved by the new bundle. However, such cases are fairly uncommon in practice because the deployment involves many nodes and the rate of producing concurrent bundles is high.
\item[Validation:]
First, $x$ validates the integrity of every bundle on the two selected paths. If any structurally invalid transaction or bundle is found, it is excluded from the Tangle.
Secondly, $x$ validates that the two paths contain no conflicting bundles that lead to double-spending. To detect possible double-spending or prevent spending a larger sum than what is available in an account, $x$ needs to keep values of all addresses in all the verified bundles on the two selected paths. If any of the values turns negative, then we know that the paths include conflicting transactions but we do not necessarily learn which transactions are conflicting.  

If the validation fails because of any of the above two issues, the tip selection algorithm is invoked again to find a different pair of tips and paths. Then, the validation is performed for the new pair of paths. This happens repeatedly until a valid pair is found. Some additional optimizations are implemented in case of structurally invalid bundles, which we do not cover in our description.

In Figure~\ref{fig:tangle-example}, no bundle violates structural integrity but bundles $D$ and $D'$ are conflicting. In this case, node 1 selects a new pair of tips as shown in Figure~\ref{fig:tangle-example-reattached}. The new pair of paths passes the validation so that the algorithm stops.

\item[Bundle propagation:]
$x$ propagates the new bundle along with the id of the two selected tips using the communication layer. Interestingly, a proof of neither correct tip selection nor performed validation is supplied in the message. When a node other than $x$ receives the bundle from the communication layer, it performs the cheaper integrity validation for that bundle. However, it does not perform an expensive detection of double-spending nor does it traverse any path from the genesis block to the new bundle.
\item[Monitoring of confirmation and reattachment:]
The validation and tip selection algorithms include a lot of randomness as we discuss below. Because of this randomness, some bundles may get confirmed much faster than the others, while some valid bundles may in fact be never confirmed and be \emph{left behind}. In view of this, $x$ needs to monitor the Tangle for signs of confirmation, which we also discuss below. IOTA introduces the concept of reattachment: node 1 may create a new bundle in the future that specifies the very same transfer as bundle $O$ in the illustration, and inject this new bundle into the Tangle by performing all the above steps. In this case, the new bundle will be conflicting with $O$ so that node 1  should only do it if $O$ is not getting confirmed for a long period of time.
\end{asparadesc}



The tip selection algorithm is essentially a weighted random walk through the Tangle starting from the genesis bundle and moving toward the tips. At each step, the algorithm proceeds from the current bundle to one of the bundles directly approving the current bundle. The probability of choosing a bundle as next step is proportional to the cumulative weight of that bundle.  The cumulative weight of each bundle $Y$ is equal to the own weight of $Y$ plus the sum of own weights of all bundles that directly or indirectly approve $Y$. Figure~\ref{fig:tangle-example} illustrates the cumulative weight of each bundle (denoted as \emph{cw} inside the bundles) as calculated by node 1 (the values for nodes 2 and 3 will be different). For instance, the cumulative weight of $A$ is equal to the sum of weights of bundles $A$, $C$, $D$, $E$, $F$, $G$, $H$, $I$, $J$, $D'$, $K$, $L$, $M$, and $N$. When the random walks determines the next step from the genesis bundle, it will make a probabilistic choice between bundles $A$ and $B$ based on their cumulative weights (of $21$ and $23$ respectively).
The random walk stops when it reaches one of the tips.



Based on the above algorithm, it is clear that choosing a bigger weight for a new bundle increases the probability of that bundle to be selected by the random walk performed during the tip selection. 
This ultimately results in probabilistically faster confirmation for a bundle with a large weight. 

Besides, it is obvious that the cumulative weight of a bundle incrementally increases as it is getting approved directly or indirectly by new bundles added to the Tangle. As a result, the cumulative weight can be used as a metric for confirming bundle inclusion in the Tangle, similarly to how the number of subsequently added blocks is used as a metric for confirming block inclusion in Bitcoin and Ethereum. This is important, e.g., when a node decides to reattach the bundle. Notably, heavy bundles tend to gain the weight faster due to the weighted random walk. In some scenarios, this may lead to situations when a bundle is never confirmed, thereby rendering the reattachment mechanism essential. However, as the number of users and transactional rate in the network increase, the probability and speed of confirming any bundle regardless of its weigh grow.

Apart from the cumulative weight, there exists another metric in IOTA used for monitoring bundle inclusion. The metric is called \emph{confirmation confidence} and denoted \emph{cc} in Figure~\ref{fig:tangle-example}. The confirmation confidence of a bundle is the percentage of the tips which directly or indirectly approve the bundle. In Figure~\ref{fig:tangle-example}, if we consider the Tangle status on node 1 before $O$ is included, the confirmation confidence of bundle $D$ is \%75. This is because among the four tips of $G$, $H$, $I$, and $N$, the three tips $G$, $H$, and $I$ are approving $D$ while $N$ is not. Note, however, that confirmation confidence is not always a good indicator of bundle inclusion if the Tangle copies on different nodes are diverging. For example, the confirmation confidence for bundle $K$ is \%100 on node 3 but this is only because node 3 does not know about bundles $G$, $H$, and $I$ yet.


In terms of Tangle security, several attacks and defenses have been considered in the literature~\cite{tangle,sok-dag}. The \emph{replay attack}~\cite{replay-blog,replay-netherlands} performs double-spending by exploiting the reattachment mechanism: the reattached bundle normally invalidates the original bundle but in the attack, both bundles remain valid, which results in double-spending. The attack only works under reuse of addresses, which is within the control of the users and which the users should strive to avoid as per IOTA recommendation. In the \emph{parasite chain attack}~\cite{tangle,parasite-chain}, the attacker aims to replace the current Tangle with its own subgraph, which the attacker builds in secret while taking steps to ensure that the secret subgraph will have sufficiently high cumulative weights. To perform double-spending, the attacker initiates two conflicting transactions. The first transaction is injected into the current Tangle while the second transaction is included into the secret subgraph. After the first transaction is confirmed and the attackers receives the services or benefits for the payment, the attacker broadcasts all bundles in the secret subgraph, which replaces the Tangle. This way the attacker gets the second transaction confirmed as well. In the \emph{splitting attack}~\cite{tangle,gal-splitting}, the attackers partitions the Tangle by making sure that two subgraphs within the Tangle maintain a similar cumulative weight over time, and injecting transactions so as to maintain this balance. Such partitioning allows the attacker to perform double-spending by placing two conflicting transactions, one in each subgraph.

The author of~\cite{tangle} proposes the following idea to defend against the parasite chain and splitting attacks. The probabilistic choice of the next step in the random walk is governed by parameter $\alpha$ in addition to the cumulative weights of candidate vertices. A big value of $\alpha$ means that the vertex with a larger cumulative weight wins with a very high probability while a small value of $\alpha$ means that the winning chance for a heavier vertex is only slightly higher. This work shows that a large value of $\alpha$ reduces the risk of successful parasite chain and splitting attacks. Unfortunately, larger values of $\alpha$ increase the probability for a bundle to be left behind and never confirmed~\cite{g-iota,parasite-chain}. In~\cite{parasite-chain}, the authors propose a detection mechanism for parasite chains, with the idea of using smaller values of $\alpha$ in a normal situation but adaptively increasing the values if a parasite chain is detected. A different idea advocated in~\cite{g-iota} is to modify the tip selection algorithm so as to explicitly search for left-behind tips and increase the chance of their selection.


Since the tip selection and validation are nontrivial algorithms in IOTA, it is interesting to consider incentives to perform them for node $x$ that wants to issue a new bundle. This is especially important because during the bundle propagation phase, $x$ does not supply any proof of correct tip selection and validation. However, while a node receiving a bundle from $x$ does not perform a random walk with validation immediately, it will do so upon creating its own bundle in the future. 
If $x$ does not perform the validation properly and approves two tips from two traversal paths that are conflicting, this may also negatively affect the probability of confirmation for the new bundle produced by $x$. In the illustration of Figure~\ref{fig:tangle-example}, if $O$ approves $G$ and $N$, then the conflict between $D$ and $D'$ may be discovered by a later random walk, which will reduce the probability of $O$ to be selected. However, there are multiple paths from the genesis bundle to $O$ that do not traverse $D$ and $D'$, in which case $O$ may still be selected. Note that if a conflict between $D$ and $D'$ is later detected by other nodes, it is impossible for them to establish whether $x$ deliberately neglected validation or simply was unlucky with detection because $x$ selected non-conflicting paths leading to $G$ and $N$ during its random walk. The exact quantification of the reduction in confirmation probability requires a very complex probabilistic analysis, which has not been performed in the literature to the best of our understanding.

In the context of tip selection, different tips have different probability to be reached by the random walk. If $x$ just arbitrarily selects two tips without performing the random walk, this may result in reduced probability for the new bundle to be reached by the random walk performed by other nodes at a later point. As a result, $x$ will be punished by a higher confirmation time. On the other hand, $x$ may minimize the confirmation time for its bundle by performing a deterministic walk which selects a vertex with the biggest cumulative weight at each step. Eliminating the randomness during the walk may exacerbate the problem of heavy vertices gaining weight very fast, resulting in a higher number of reattachments. There exist considerations, however, that detract from the benefits of such a selfish node strategy, as outlined in~\cite{tangle}.



The consensus algorithm in IOTA is relatively lightweight and scalable in the sense that it can sustain a higher rate of adding new transactions to the storage and confirming them. This improvement eliminates the need in two design elements present in the other three systems we consider: batching transactions into a block and having long intervals between consecutive blockchain updates. 
As a result of allowing shorter intervals between updates, PoW is much more lightweight in IOTA compared to Bitcoin and even Ethereum. 
While the typical IOTA throughput has been reported as 45-50 transactions per second~\cite{iotaTps}, IOTA has mentioned a throughput of above 500 transactions per second achieved under a proprietary stress test~\cite{iotaTps1}.
These advantages do come at a cost. Since discovery of conflicts is probabilistic, it may take a lot of time until double-spending is discovered. The Tangle storage requires more space compared to systems that use a chain of blocks, as explained in Section~\ref{sec:storage-org}. Most importantly, the paths produced by the tip selection algorithm become long as the Tangle grows, resulting in a very expensive verification procedure. 



To address the challenge of scaling the Tangle and the validation algorithm, IOTA employs a snapshot process that prunes the transaction history (see Section~\ref{sec:retention}) and shortens validation paths by creating a new snapshot bundle that effectively serves as a new genesis bundle. The snapshot process is periodically performed by the IOTA Foundation, which goes against the blockchain spirit of eliminating the trust in any single entity or organization. The Coordicide project~\cite{coordicide} is focused on removal of coordination and in particular, the need for a snapshot process.




\subsubsection{Defense against DoS attacks}
\label{sec:dos}

DoS attacks in Bitcoin, Ethereum and IOTA are mitigated by the Proof-of-Work mechanism. As discussed in Section~\ref{sec:bitcoin}, a PoW algorithm requires solving cryptographic puzzles for 
creating new blocks.
If a message with a new block or transaction does not contain a valid solution to a cryptopuzzle, the message is rejected without further processing, which makes it easier for the system to filter out spurious messages. 

However, PoW requires a significant amount of energy~\cite{energy} and wasted computation performed by legitimate miners. 
Proof-of-Stake (PoS) protocols were developed as energy-saving alternatives to PoW~\cite{pos}. Instead of solving a cryptopuzzle as in the PoW, a block creator is selected based on its stakes (the number of digital tokens that it holds) in PoS. As there is no heavy computation process in PoS, the block creation time is much shorter than in PoW, thereby resulting in a higher transaction throughput.
In order to prevent DoS, a block proposer needs to make a deposit to gain the right to make a proposal. Since the deposit is locked until the next block is selected, this makes it more expensive for the miner to create multiple proposals within a short period of time. Besides, the deposit can be confiscated for malicious behaviors such as performing a DoS attack.


The Ethereum network has developed the Casper protocol~\cite{casper} in an attempt to ease the transition from the current PoW protocol to a pure PoS protocol. Ethereum 2.0 will deploy the Casper protocol on top of the existing PoW protocol, prior to switching to a pure PoS protocol in future releases. So, while blocks are still going to be mined via PoW in Ethereum 2.0, every fixed interval (every 50 blocks~\cite{checkpoint}) is going to be a PoS checkpoint where the finality of blocks is assessed by a dynamic committee that votes via a BFT protocol.
To join the committee, a validator has to make a deposit to gain a voting right proportional to that deposit. 


An additional vector of attacks for a user in Ethereum is to inject a computationally- or storage-expensive transaction that would consume resources on all of the full nodes in the network.  The gas budget concept (see Section~\ref{sec:ethereum}) plays an essential role in defending against this attack: the attacker itself would need to pay a substantial amount of gas in order to launch the attack.



While IOTA is using PoW as well, the cryptopuzzles in IOTA are much simpler compared to Bitcoin and Ethereum which leads to a higher throughput. The reason for choosing a lower-difficulty PoW in IOTA is that IOTA uses PoW just as a spam protection~\cite{coordicide}; whereas, in Bitcoin and Ethereum, PoW is additionally used for controlling the rate of block production. On the other hand, a separate PoW is required for each transaction in a bundle since a bundle may theoretically contain an unlimited number of transactions. While the current implementation of IOTA is utilizing a simple PoW mechanism, the IOTA Coordicide~\cite{coordicide} project has proposed an adaptive rate control mechanism which intelligently adjusts the difficulty of the PoW per IOTA node based on different factors, such as a number of recently-issued transactions and reputation of the issuer~\cite{coordicide}. 


Hyperledger Fabric does not have a single main deployment. Instead, there is a multitude of proprietary deployments, where each deployment is partitioned by channels (see Section~\ref{sec:communication}). Besides, the system is permissioned and requires authorization for joining each channel. These factors significantly mitigate the risk of DoS attacks.

\subsubsection{Querying blockchain}
\label{sec:querying}


In Section~\ref{sec:datastorage}, we describe different data items stored by the data recorders of each blockchain system. Any data stored by the recorders can in principle be locally queried by them. If a data recorder acts as a query responder, it can essentially answer any queries related to the stored data that are initiated by query issuers.

However, the fact that the data is stored does not necessarily mean that there exist means to query it efficiently. The functionality of blockchain systems is different compared to general-purpose databases (see Section~\ref{sec:reflection}). The native implementation of most blockchain systems only efficiently supports limited queries as simple as retrieving a data item by its hash code. The main reason for the limited support of queries is the limitations in the default database implementation of the blockchain systems. For example, levelDB used in Bitcoin, Ethereum, and Hyperledger Fabric as a key-value database supports only querying based on keys. Another reason is the way of utilizing the database for storing and accessing data. For example, retrieving values in Ethereum requires traversing multiple trie structures (see Figure~\ref{fig:disk} in Section~\ref{sec:on-disk}), and accessing values associated with the queried hash code (i.e. the key) requires searching multiple keys over the disk, which is time-consuming~\cite{etherql}.
 

Due to these reasons, supporting analytical and complex queries (querying based on values other than keys) such as filter, aggregation, and sorting queries on the blockchain systems require some extra considerations and functional components.
The modularity of Hyperledger Fabric and its focus on highly configurable proprietary deployments allow for modifying the native implementation. Specifically, replacing the default database module of LevelDB with CouchDB allows for supporting more complex search semantics in an efficient way~\cite{couchdb}. For example, as CouchDB allows storing data in JSON format, it is possible to query by data values, and not just by keys. The tradeoff, however, is that CouchDB requires more disk space for storing the data compared to LevelDB.

Research proposals of adding an extra query layer to existing blockchain systems have also emerged. For example, \emph{EtherQL}~\cite{etherql} supports a set of analytical queries such as aggregation and top-k queries on top of Ethereum. These proposals, however, have not been implemented in the actual systems as of 2021.


The most common way of solving the issue of rich query semantics is designating a subset of data recorders as dedicated query responders (see Section~\ref{sec:participants}).  In practice, query responders maintain a separate database in addition to the blockchain data, which allows them to answer richer queries efficiently. Such designated query responders are used in Bitcoin, Ethereum, and IOTA. \emph{Blockchain.com}~\cite{bitcoinquery} and \emph{etherscan.io}~\cite{etherscan} are websites that allow anyone to search for a specific address, transaction, or block on Bitcoin and Ethereum databases respectively. 
However, these sites do not provide more advanced search functionalities such as looking for transactions with a specific amount of input.
Some other services store the blockchain in an SQL database and provide the ability to issue general SQL queries. Notably, \emph{BlockchainSQL}~\cite{blockchainsql} does it for Bitcoin while \emph{Anyblock}~\cite{anyblock} provides such a service for Ethereum and some other blockchain systems. 
%
IOTA foundation provides a website (\emph{explorer.iota.org}~\cite{iota-explorer}) as an external dedicated service for querying IOTA. This website supports basic search functionalities for the IOTA Tangle based on keys, such as searching based on a transaction hash, a bundle hash, or an IOTA address. As an example, searching for an IOTA address through this website gives information about the balance of the address and all transactions related to that address. \emph{thetangle.org}~\cite{tangle-org} is another external dedicated service for IOTA, which provides visualization for a number of specific queries.

\begin{table*}
	\setlength\tabcolsep{3pt} \setcellgapes{3pt} \makegapedcells
	\centering
	\caption{Contract layer of different blockchain systems}
	\label{tab:contract}
	\scalebox{1}{
		\begin{tabular}{|Y{3.1cm}|Y{2.2cm}|Y{3.8cm}|Y{4.9cm}|Y{2.1cm}| @{}}	
				\hline
				\textbf{Features} & \textbf{Bitcoin} & \textbf{Ethereum} & \textbf{Hyperledger Fabric} & \textbf{IOTA} \\
				\hline
				Type & Scripting system & Smart contracts & Smart contracts packaged into a chaincode & No support yet (in progress) \\   
				\hline
				Programming language & Forth-like & Several languages, Solidity as the most popular one & Go, Node.js, Java & - \\
				\hline
				Turing completeness & No & Yes & Yes &  - \\
				\hline
				Executing computing devices & Full nodes & Full nodes & Specified by endorsement policy & - \\ 
				\hline
				Execution language & Bitcoin script & Ethereum Bytecode & Programming language-dependent & - \\
				\hline
				Execution environment & Script interpreter	& Ethereum Virtual \newline Machine  & Docker container & - \\
				\hline
		\end{tabular}}
		\vspace{-5mm}
	\end{table*}

However, such designated query responders pose two major challenges. First, the exact rich query semantics is up to a specific implementation; it is not standardized in any way. Secondly and most importantly, they do not provide the security, dependability, and availability guarantees commonly attributed to blockchain systems. 
This is because they are managed by individual computing devices that are fundamentally untrusted in the blockchain paradigm. In particular, query responders can decide on an access control policy that is not aligned with the access control policy of a particular permissioned or permissionless blockchain system. For example, \emph{Anyblock.tools} provides different query services based on the paid subscription tier of the users. 

In summary, query responders serve two purposes: (a) to provide rich query semantics otherwise unavailable in the blockchain system and (b) to provide information to computing devices that do not belong to the blockchain network.

\subsection{Contract Layer}
  \label{sec:contract}

The contract layer allows users of the blockchain system to develop programmatic extensions and install them atop blockchain. Smart contracts are similar to distributed objects in nature. 
A contract implements a collection of procedures, each of which can be invoked remotely and executed as a transaction. A contract may additionally invoke transactions implemented by other contracts. 

Features of the different blockchain systems in the contract layer are shown in Table~\ref{tab:contract}. As Bitcoin was designed with a specific defined application in mind, it does not implement a general-purpose contract layer. Instead, it uses a simple scripting system for its transactions. Each transaction contains an output and an input script~\cite{bitcoinText}. When a transaction executes, its inputs are verified. Let us assume transaction $T_1$ has an input $I$, which is an output of transaction $T_0$. In this case, $I$ is verified by executing the output script of $T_0$ followed by an input script of $T_1$. 
By combining the input script of the new transaction with the output script of the referenced transaction, a full node can validate that the new transaction can redeem the previous transaction output. The reason for concatenating the input and output scripts is that the output script specifies the verification procedure, whereas the input script contains cryptographically secure information about the redeeming node (specifically, the public key and a signature of that public key), which is used as an input to the verification procedure. Bitcoin provides a default implementation of an output script (called scriptPubKey) and a default implementation of an input script (called scriptSig). This default implementation tests if the hash of the provided public key matches the hash in the scriptPubKey, and then checks the signature against the public key to finally validate the new transaction.
Bitcoin full nodes are responsible for validating the transactions by executing Bitcoin scripts in their script interpreter environment. The programming language of Bitcoin scripts resembles Forth~\cite{forth}. Similar to Forth, Bitcoin scripts are stack-based and they use reverse-Polish notation (processed from left to right). Since Bitcoin scripts are not intended for writing general-purpose applications, they are not Turing-complete~\cite{bitcoinText}. In particular, the language does not support jumps and loops so that malicious users cannot create code that would waste the power and computing resources of the Bitcoin network. 

Instead of simple scripts, Ethereum utilizes smart contracts, which can be used for implementing a variety of decentralized applications.  One of the main differences between the Ethereum smart contracts and Bitcoin scripts is that smart contracts are Turing complete, which is essential for a general-purpose blockchain platform.
When a contract procedure is invoked, it is executed on top of the blockchain platform by Ethereum full nodes, using the Ethereum Virtual Machine (EVM) environment. 
As discussed in the storage layer Section~\ref{sec:datastorage}, a smart contract stores its own data.
Remote invocations that update the contract state are executed under transactional semantics.

Currently, the most common and supported language for writing smart contracts is Solidity~\cite{solidity,solidity1}. What makes this language popular is that Solidity is object-oriented and high-level, and it provides a rich support for developing contracts. Solidity has similarities to C++, Python, and JavaScript. LLL (a low-level Lisp-like language)~\cite{lll} and Vyper (Python-derived)~\cite{vyper} are other popular alternative languages for writing smart contracts. 
A code written in any of these languages needs to be complied to Ethereum Bytecode in order to run in the EVM environment.

Hyperledger Fabric also has the concept of smart contracts, similar to that of Ethereum. However, related smart contracts can be grouped together and packaged into a chaincode in Fabric; While smart contracts are governing transactions, chaincode governs how smart contracts are packaged for deployment~\cite{contractsHyperledger}. When a chaincode is instantiated on a channel, an endorsement policy is defined for it~\cite{contractsHyperledger}. Therefore, the endorsement policy applies to all smart contracts defined whithin the same chaincode in the context of a channel. In Fabric, unlike Bitcoin and Ethereum, transactions and smart contracts are only executed by endorsers (see Sections~\ref{sec:manipulation-fabric} and~\ref{sec:hyperledger}), which create W-R sets. The other peers validating or applying the transaction only process W-R sets instead of executing the code of the transaction.
Currently, smart contracts of Fabric can be written in Go, Node.js, and Java~\cite{chaincode}, which are all Turing complete languages. Endorsers execute the chaincode and its smart contracts in a Docker container environment corresponding to the contract programming language~\cite{chaincode}. Docker container is an isolated sandbox inside the endorser whose purpose is to prevent the chaincode from accessing the endorser's local environment and resources. 

The IOTA foundation has stated that smart contracts will not be a feature of IOTA core~\cite{iotaContract}. However, there are some projects such as Qubic protocol~\cite{qubic} which aim to design a contract layer on top of IOTA Tangle in order to deploy smart contracts and other applications on it.

 



\section{Related Work}
\label{sec:relatedWorks}

All published surveys on blockchain technology can be categorized as surveys on individual aspects or on blockchain as a whole. 
For the former survey type, \cite{Marko2}, \cite{natoli2019deconstructing}, \cite{xiao2020survey}, and \cite{gupta2021fault} concentrate specifically on the consensus mechanisms of blockchain.  In \cite{zhang2019security}, \cite{salman2018security}, \cite{saad2020exploring}, and \cite{homoliak2020security}, the focus is on the security analysis. Smart contracts are covered in depth in \cite{bartoletti2017empirical} and \cite{varela2021smart}. Data management and analysis of blockchain data are investigated in \cite{eberhardt2017or} and \cite{vo2018research}. Network layer aspects of permissionless blockchains are addressed in \cite{neudecker2018network}. Surveys of \cite{yang2019integrated}, \cite{liu2020blockchain}, \cite{sharma2020blockchain}, \cite{gai2020blockchain}, and \cite{nguyen2020integration} review integration of blockchain with other technology, namely with edge computing, machine learning, cloud storage, cloud computing, and cloud of things respectively. The works of~\cite{lao2020survey}, \cite{ali2018applications}, \cite{de2020survey}, and \cite{xie2019survey} survey on applying blockchain to various application domains such as IoT, healthcare, and smart cities.

\newcommand{\weaktext}[1]{{\color{lightgray} #1}}

\begin{table*}
	 \setlength\tabcolsep{3pt} \setcellgapes{3pt} \makegapedcells
	\centering
	\caption{Comparison with state of the art surveys on blockchains as a whole}
	\label{tab:relatedWorks}
	\scalebox{0.93}{
	\begin{tabular}{|C{0.3cm}|C{1.9cm}|C{1.9cm}|*{14}{C{0.8cm}|}}
	 \hline
	 \multicolumn{3}{|C{4.1cm}|}{Main analayzed aspects of blockchains}& Our work & \cite{bonneau2015sok} & \cite{tschorsch2016bitcoin} & \cite{taxonomy} & \cite{kolb2020core}& \cite{haffke2017technical} &  \cite{2017ontology} & \cite{tasca2017ontology} & \cite{valenta2017comparison} & \cite{sok-dag} & \cite{untangling} & \cite{kannengiesser2020trade} & \cite{ellervee2017comprehensive} & \cite{belotti2019vademecum} \\
	 \hline
	 
	 1 & \multicolumn{2}{C{3.8cm}|}{Comparison of alternative definitions}& \cmark\textsuperscript{1} & \xmark\textsuperscript{2} & \xmark & \xmark & \xmark & \xmark & \xmark & \xmark & \xmark & \xmark & \xmark & \xmark & \xmark & \xmark \\
	 \hline
	 
	 2 & \multicolumn{2}{C{3.8cm}|}{Roles and entities}& U\textsuperscript{3} & \xmark & \xmark & \xmark & \xmark &  \xmark & \xmark & \xmark & \xmark & \xmark & \xmark & \xmark & S\textsuperscript{4} & U \\
 	 \hline
 	 
 	 3 & \multicolumn{2}{C{3.8cm}|}{Hardware layer}& \makecell{B\textsuperscript{5},E\textsuperscript{6}, \\ H\textsuperscript{7},I\textsuperscript{8}} & B,etc. & B,etc. &  \xmark & \xmark & \makecell{B,E,\\etc.}  & \xmark & \xmark & \xmark & \xmark & \xmark & \xmark & \xmark & \weaktext{B}\textsuperscript{9},\weaktext{E}\textsuperscript{10} \\
 	 \hline
 	 
 	 \multirow{3}{*}{4} & \multirow{3}{*}{\makecell{Data storage\\ layer}} & State tracking & \makecell{B,E,\\H,I} & B & B & \xmark & \xmark & \makecell{B,E,\\etc.} & \xmark & \xmark & \xmark & \weaktext{I}\textsuperscript{11},etc. & \xmark & \xmark & \makecell{\weaktext{B},\weaktext{E},\\etc.} & \makecell{B,E,\\H,etc.} \\
 	 \cdashline{3-17}
 	 & & Disk & \makecell{B,E,\\H,I} & \xmark & \weaktext{B} & \xmark & \xmark & \xmark & \xmark & \xmark & \xmark & \xmark & \xmark & \xmark & \xmark & \weaktext{H}\textsuperscript{12} \\
 	 \cdashline{3-17}
 	 & & Memory & \makecell{B,E,\\H,I} & \xmark & \xmark & \xmark & \xmark & \xmark & \xmark & \xmark & \xmark & \xmark & \xmark & \xmark & \xmark & \xmark \\
 	 \hline
 	 
  	 5 & \multicolumn{2}{C{3.8cm}|}{Communication layer}& \makecell{B,E,\\H,I} & B & B & \xmark & \xmark & \makecell{B,\weaktext{E},\\etc.}  & \xmark & \xmark & \xmark & \xmark & \xmark & \xmark & \makecell{\weaktext{B},\weaktext{E},\\etc.} & \makecell{\weaktext{B},\weaktext{E},\\\weaktext{H},etc.} \\
 	 \hline
 	 
  	 \multirow{4}{*}{6} & \multirow{4}{*}{\makecell{Data \\ manipulation\\ layer}} & Consensus algorithm & \makecell{B,E,\\H,I} & B & \makecell{B,etc.} & \makecell{\weaktext{B},\weaktext{E},\\etc.} & \makecell{B,\weaktext{E},\\etc.} & \makecell{B,\weaktext{E},\\etc.} & \xmark & \makecell{\weaktext{B},\weaktext{E},\\etc.} & \makecell{\weaktext{E},\weaktext{H},\\etc.}. & I,etc. & \xmark & \makecell{\weaktext{B},\weaktext{E},\weaktext{H},\\ \weaktext{I},etc.} & \makecell{\weaktext{B},\weaktext{E},\\etc.} & \makecell{\weaktext{B},\weaktext{E},\\ \weaktext{H},etc.} \\
 	 \cdashline{3-17}
 	 & & Quantitative performance comparison & \makecell{B,E,\\H,I} & \xmark & \xmark & \xmark  & \xmark &  \xmark & \xmark & \xmark & \xmark & \xmark & \makecell{E,H,\\etc.} & \xmark & \xmark & \xmark \\
 	 \cdashline{3-17}
  	 & & DoS prevention & \makecell{B,E,\\H,I} & B,etc. & B,etc. & \xmark & \xmark & \xmark & \xmark & \xmark & \xmark & \xmark & \xmark & E,\weaktext{H} & \xmark & \weaktext{B},\weaktext{E} \\
 	\cdashline{3-17}
 	& & Querying & \makecell{B,E,\\H,I} & \xmark & \xmark & \xmark & \xmark & \xmark & \xmark & \xmark & \xmark & \xmark & \xmark & \xmark & \xmark & \xmark \\
 	 \hline
 	 
   	 7 & \multicolumn{2}{C{3.8cm}|}{Contract layer}& \makecell{B,E,\\H,I} & B & B,etc. & \weaktext{B},\weaktext{E} & \makecell{E,\weaktext{H},\\etc.} & B,E & \xmark & \makecell{\weaktext{B},\weaktext{E}} & \makecell{\weaktext{E},\weaktext{H},\\etc.} & \xmark & \makecell{B,E,H,\\I,etc.}  & \makecell{B,E,\\etc.}  & \xmark & \makecell{B,E,\\H,etc.} \\
 	 \hline
\end{tabular}}
	\footnotesize{\textsuperscript{1}\cmark: Considered, \textsuperscript{2}\xmark: Not considered, \textsuperscript{3}U: Universal, \textsuperscript{4}S: System specific, \textsuperscript{5}B: Bitcoin, \textsuperscript{6}E: Ethereum, \textsuperscript{7}H: Fabric, \textsuperscript{8}I: IOTA, \textsuperscript{9}\weaktext{B}: Briefly mentioned in Bitcoin, \textsuperscript{10}\weaktext{E}: Briefly mentioned in Ethereum, \textsuperscript{11}\weaktext{I}: Briefly mentioned in IOTA, \textsuperscript{12}\weaktext{H}: Briefly mentioned in Fabric}
\end{table*}


In contrast, blockchain surveys of the second type provide a general exposition of blockchain systems. However, the coverage significantly varies across specific surveys. 
As our paper falls in this category of blockchain surveys, we explore the related surveys of this type in greater detail and compare them with our work in the rest of this section (see Table~\ref{tab:relatedWorks}).

The study of~\cite{bonneau2015sok} has been one of the first surveys on blockchains as a whole. However, it is limited to the cryptocurrency use case of blockchain. This work reviews research perspectives and challenges for Bitcoin and other cryptocurrencies. It focuses on Bitcoin and presents transactions (including scripts), the consensus protocol, and the communication network as the main technical components. Incentivizing correct behavior in Bitcoin, transition towards more powerful and energy-efficient customized hardware, and security issues of cryptocurrencies are the other important aspects covered in \cite{bonneau2015sok}.    
 Another technical survey on Bitcoin and cryptocurrencies~\cite{tschorsch2016bitcoin} is similar to \cite{bonneau2015sok} in terms of the goals.
Additionally, it includes a tutorial-style introductory part and an in-depth overview of more recent existing literature.  

The authors of~\cite{taxonomy} compare Bitcoin with Ethereum and propose a design taxonomy based on this comparison.
Discussion of the architectural design in~\cite{taxonomy} is structured by topics of decentralization, computation, infrastructural configuration, among others. Based on this discussion, the work evaluates consensus protocols and system performance. Bitcoin scripts and Ethereum smart contracts are also compared as another principal aspect of blockchain in this paper. Similar to~\cite{taxonomy}, \cite{kolb2020core} is another survey using Ethereum and Bitcoin as a case study to describe the inner workings of blockchain in detail. The survey covers blockchain incentive structures, smart contracts, and scalability issues such as transaction throughput and latency. 
The study of~\cite{haffke2017technical} provides yet another comparison of Bitcoin with Ethereum which considers a number of additional aspects: UTXO and account model for tracking the states, different hardware used for mining in Bitcoin and Ethereum, and the differences in the network layer. Similarly to \cite{taxonomy} and \cite{kolb2020core}, consensus protocol and scripting language of Bitcoin and Ethereum are also reviewed in~\cite{haffke2017technical}.
As another work that compares Bitcoin and Ethereum, \cite{gervais2016security} introduces a quantitative framework to analyze the security and performance implications of various consensus and network parameters of PoW blockchains. Impact analysis of parameter choices such as block interval, stale block rate, and average block size on the network propagation time and the throughput of PoW blockchain systems is one of the main contributions in~\cite{gervais2016security}.      

Lack of formalization and standardization in blockchain technology has prompted a research on ontologies that aim at providing a vocabulary of key blockchain terms. A high-level ontology for concepts and definitions without technical details is proposed in~\cite{2017ontology}. 
A more technical ontology is introduced in \cite{tasca2017ontology}.
Furthermore, this work provides an overview of a number of blockchain components with the main focus on Bitcoin and Ethereum, namely of consensus, transaction model, scalability limits, scripting language, and the reward incentives. 
These ontologies, however, do not survey alternative term definitions in this turbulent area and do not aim to analyze the differences.

In contrast to the other related works that concentrate on Bitcoin as a basis for analyzing blockchain design, \cite{valenta2017comparison} focuses on aspects unrelated to cryptocurrency.
This paper compares Ethereum, Hyperledger Fabric, and Corda in terms of participation of peers, consensus, and smart contracts to show the most suitable use cases for each blockchain system. The only other related survey that is not based on Bitcoin is \cite{sok-dag} which provides a comprehensive analysis of DAG-based blockchain systems. In this recent survey, consensus over DAGs, performance analysis, and transaction models are considered based on over 20 DAG-based systems including IOTA.

The work of~\cite{untangling} proposes a benchmark framework for evaluating the performance of a variety of blockchain networks, both public and private. The authors apply the framework to Ethereum, Hyperledger Fabric, and another private blockchain system and provide both qualitative and quantitative performance analysis. Additionally, the authors focus on smart contracts and provide a short survey of other blockchain-related mechanisms without specifically focusing on individual systems.


The comprehensive survey of~\cite{kannengiesser2020trade}  identifies a list of 40 DLT characteristics that are fundamental for assessing the suitability of DLT designs for applications on DLT. These characteristics are grouped into 6 categories. Then, based on the introduced characteristics, \cite{kannengiesser2020trade} proposes 24 trade-offs in the DLT design. Performance is one of the categories analyzed in this survey. As a consequence, performance characteristics such as block creation interval, block size limit, throughput, etc. are explored for Bitcoin, Ethereum, Hyperledger Fabric, IOTA, and other blockchains. According to the classification of~\cite{kannengiesser2020trade}, smart contracts are one of the characteristics mentioned for the flexibility category; thus, they are discussed in the context of a trade-off with the other DLT characteristics in the survey. In addition to the characteristics explored in~\cite{kannengiesser2020trade}, consensus mechanisms of Bitcoin, Ethereum, Hyperledger Fabric, and IOTA are briefly introduced, and DoS prevention solutions of Ethereum and Hyperledger Fabric are explained.  

Data models, network discovery process, and consensus process across four blockchain platforms including Bitcoin and Ethereum are compared in~\cite{ellervee2017comprehensive}. This work also briefly considers actors and roles in each of the systems. The excellent recent tutorial of~\cite{belotti2019vademecum} touches upon a large number of aspects and state-of-the-art mechanisms related to blockchain, including roles, entities, data model, communication protocols, consensus mechanisms, smart contracts, and much more. The tutorial mentions the implementation of as many as seven systems, including Bitcoin, Ethereum, and Hyperledger. The main goal of \cite{belotti2019vademecum} is to provide a very broad picture of the state-of-the-art and answer a number of important high-level questions, while we provide in-depth design comparison of the four systems.


In Table~\ref{tab:relatedWorks}, we explain how our survey differs from the aforementioned studies. The rows of the table correspond to seven major aspects of blockchain:
(1) comparison of alternative definitions, (2) taxonomy of roles and entities, (3) coverage of the hardware layer, (4) coverage of the data storage layer including state tracking, on-disk, and in-memory storage, (5) coverage of the communication layer, (6) coverage of the data manipulation layer at fine granularity including consensus algorithm, quantitative performance comparison, DoS prevention, and blockchain querying, and (7) coverage of the contract layer. 
The coverage of aspects 3 to 7 is shown in the table at the granularity of individual systems we consider in our survey.

The first aspect, i.e, the comparison of alternative blockchain definitions, has not been considered in the state-of-the-art surveys and is one of the contributions of our work. In Section~\ref{sec:reflection} we identify five blockchain definitions used in the literature and existing blockchain-based systems. The second aspect is a  comprehensive taxonomy of roles and entities. While there are works that consider roles and entities in specific systems, the only prior cross-system taxonomy is given in~\cite{belotti2019vademecum}  to the best of our knowledge. We extend this taxonomy to additional roles and present the implementation of each system in the light of this taxonomy. Our analysis of roles and entities, given in Section~\ref{sec:participants}, is universal and applicable to all blockchain systems. 

The third aspect for comparison is coverage of the hardware layer. While a few related works discuss the mining devices of Bitcoin or Ethereum, we compare all four systems in terms of the limiting resources, cryptopuzzle solving device, etc. We also consider how additional hardware is used for security.

Fourth, we provide a comprehensive coverage of the data storage layer in Section~\ref{sec:datastorage} including state tracking, on-disk and in-memory storage. While other surveys have considered the storage in Bitcoin, in-depth understanding of the storage in Ethereum, Hyperledger Fabric, and IOTA requires reading system documentation, blog posts, and even the source code. To the best of our knowledge, we provide the first in-depth survey coverage for these systems.


The fifth aspect is coverage of the communication layer. While the basic communication protocol in Bitcion is explained in a few related works, our description in Section~\ref{sec:communication} considers ordering guarantees, privacy and security, propagation time, initial peer discovery, and geographical proximity between network participants in Bitcoin, Ethereum, Hyperledger Fabric, and IOTA.


The sixth aspect of consideration is coverage of the data manipulation layer. This is possibly the most substantial layer of blockchain; it is touched upon by almost all the related works but the breadth and depth of coverage varies. Consideration of the consensus algorithm has received a lot of attention in each of the systems, yet cross-system comparisons are rare beyond the general comparison between permissioned and permissionless systems. For example, our survey is the first to provide a comprehensive comparison between consensus in Bitcoin and Ethereum and to contrast the consensus in IOTA with other systems not based on DAG. In particular, we provide the most complete presentation of the consensus protocol in IOTA that covers incentive-based attacks. 
Some of the elements pertaining to the data manipulation layer, such as DoS prevention and querying blockchain are covered in much greater detail in our survey compared to related work. When it comes to performance comparison, we contrast quantitative and qualitative findings about each individual system produced by the developers and researchers and attempt to place them in a unifying framework.


Regarding the seventh aspect, i.e., coverage of the contract layer, state-of-the-art mostly considers the scripting language of Bitcoin and smart contract languages of Ethereum. In comparison, we compare different systems by the contract executing computing devices, programming vs. execution language, and contract execution environment.


\section{Conclusions}
\label{sec:conclusions}

We have presented a comparative study of Bitcoin, Ethereum, Hyperledger Fabric, and IOTA. The study is organized by roles of the participants, system entities, as well as system layers in a cross-system blockchain architecture: hardware, storage, communication, data manipulation, and contract. We have also discussed the performance of the four systems based on the previously published information. The study has emphasized the differences in the design goals and principles between the systems. We hope that the study can be used as educational material in courses and tutorials. It is our conjecture that this first cross-system comparison will facilitate similar studies in the future, and that such studies will collectively contribute toward standardization of the area.

\section*{Acknowledgment}
We are thankful to many researchers who proofread an earlier version of this survey and provided insightful comments. A complete list of acknowledgements will be provided in subsequent versions.

\ifCLASSOPTIONcaptionsoff
  \newpage
\fi



\bibliographystyle{IEEEtran}
\bibliography{IEEEabrv,ref}
\end{document}